\documentclass[a4paper,prd,nofootinbib,preprintnumbers,reprint,twocolumn]{revtex4-1}

\usepackage{amsmath}
\usepackage{amssymb}
\usepackage{graphicx}
\usepackage{hyperref}
\usepackage{multirow}
\usepackage{natbib}
\usepackage{slashed}
\usepackage{siunitx}
\usepackage{xcolor}

\newcommand{\wilson}[2][]{\mathcal{C}^{#1}_{#2}}
\newcommand{\op}[1]{\mathcal{O}_{#1}}

\newcommand{\krf}{|\vec{k}_{\rm RF}|}

\newcommand{\MeV}{\text{MeV}}
\newcommand{\GeV}{\text{GeV}}
\newcommand{\Jhat}{\hat{J}}
\newcommand{\Vubeff}{V_{ub}^\text{eff}}
\renewcommand{\Im}[1]{{\rm Im}\left\lbrace{#1}\right\rbrace}
\renewcommand{\Re}[1]{{\rm Re}\left\lbrace{#1}\right\rbrace}
\DeclareMathOperator{\sign}{sgn}
\renewcommand{\[}{\big[}
\renewcommand{\]}{\big]}

\newcommand{\dd}{{\rm d}}

\newcommand{\refeq}[1]{eq. (\ref{eq:#1})}

\newcommand{\reffig}[1]{figure \ref{fig:#1}}
\newcommand{\refapp}[1]{appendix \ref{app:#1}}
\newcommand{\refsec}[1]{section \ref{sec:#1}}
\newcommand{\reftab}[1]{table \ref{tab:#1}}

\newcommand{\eps}{\varepsilon}
\newcommand{\para}{\parallel}

\newcommand{\eg}{{\cal e.g.\ }}

\newcommand{\ie}{{\cal i.e.}}

\newcommand\checked\checkmark

\allowdisplaybreaks

\begin{document}

\title{Analyzing $b\to u$ transitions in semileptonic $\bar{B}_s \to K^{*+}(\to K \pi)\ell^-\bar\nu_\ell$ decays}
\author{Thorsten Feldmann}
\email{thorsten.feldmann@uni-siegen.de}
\author{Bastian M\"uller}
\email{mueller@physik.uni-siegen.de}
\author{Danny van Dyk}
\email{vandyk@physik.uni-siegen.de}
\affiliation{Theoretische Physik 1, Naturwissenschaftlich-Technische Fakult\"at,
Universit\"at Siegen, Walter-Flex-Stra\ss{}e 3, D-57068 Siegen, Germany}

\date{v1}

\preprint{SI-HEP-2015-11,QFET-2015-12,EOS-2015-01}

\begin{abstract}
We study the semileptonic decay $\bar{B}_s \to K^{*+} \ell^-\bar\nu_\ell$, 
which is induced by $b\to u \ell^- \bar\nu_\ell$ transitions
at the quark level.
We take into account the standard model (SM) operator from $W$-boson exchange
as well as possible extensions from physics beyond the SM.
The secondary decay $K^{*+}\to K\pi$ can be used to 
study a number of angular observables, which are worked out in terms of
short-distance Wilson coefficients and hadronic form factors.
Our analysis allows for an independent
extraction of the Cabibbo-Kobayashi-Maskawa matrix element 
$|V_{ub}|$ and for the determination of certain ratios 
of $\bar{B}_s\to K^*$ form factors. 
Moreover, a future precision measurement of the forward-backward asymmetry in the 
$\bar{B}_s \to K^{*+} \ell^-\bar\nu_\ell$ decay can be used to 
unambiguously verify the left-handed nature of the transition operator
as predicted by the SM.
We provide numerical estimates for the relevant
angular observables and the resulting decay distributions on the basis of
available form-factor information from lattice and sum-rule estimates. 
In addition, we pay particular attention to suitable
combinations of angular observables in the decays $\bar{B}_s \to
K^{*+}(\to K\pi)\ell^-\bar\nu_\ell$ and $\bar{B} \to K^{*0}(\to
K\pi)\ell^+\ell^-$, and find that they provide complementary
constraints on the relevant $b\to s$ short-distance coefficients.
As a by-product, we perform a SM fit on the basis of
selected experimental decay rates and hadronic input functions,
which results in $|V_{ub}| = (4.07 \pm 0.20) \cdot 10^{-3}$.
\end{abstract}

\maketitle

\section{Introduction}

The value of $|V_{ub}|$ represents one of the least well-measured parameters 
in the Cabibbo-Kobayashi-Maskawa (CKM) matrix of the standard model (SM).
Moreover, at present, its inclusive determination from  $B \to X_u\ell\nu_\ell$ decays
and the extraction from 
exclusive semileptonic or leptonic decay modes lead to somewhat different 
results (see e.g.\ the review in \cite{Kowalewski:2014PDG}).
Independent phenomenological information
on $b \to u$ transitions will clearly help to better understand 
the origin of these discrepancies and the underlying theoretical 
uncertainties.
As the solution to this $|V_{ub}|$ puzzle might also be related to 
physics beyond the SM, one should also take into account possible
new physics (NP) effects;
see \cite{Buras:2010pz,Dutta:2013qaa,Bernlochner:2014ova} for recent work in
that direction. 

The proliferation of unknown parameters, which arises in 
a model-independent approach with generic dimension-6 operators
in the effective Hamiltonian, can be handled with a sufficient number
of independent experimental observables in $b \to u$ transitions.
An example is  $B\to \rho(\to \pi\pi)\ell\nu_\ell$ where the analysis of the  secondary
$\rho\to\pi\pi$ decay introduces a large number of angular
observables with different sensitivities to the individual short-distance
coefficients \cite{Bernlochner:2014ova}. This is similar to what
has been extensively used in the analysis of rare exclusive $b \to s\ell^+\ell^-$ transitions 
\cite{Kruger:1999xa,Kruger:2005ep,Egede:2008uy,Altmannshofer:2008dz,Bobeth:2012vn,Boer:2014kda}.  
Because of the large hadronic width of the $\rho$ resonance
and the question of the S- and P-wave composition of the experimentally 
measured dipion final state, a precision determination of $|V_{ub}|$ from this decay
also requires a better theoretical understanding of  
the $B \to \pi\pi\ell\nu_\ell$ decay spectrum \cite{Faller:2013dwa,Kang:2013jaa}. \\

In this article, we focus on the decay $\bar{B}_s \to K^{*+}(\to
K\pi)\ell^-\bar\nu_\ell$, which provides similar
insight into the short-distance couplings 
as the decay $B  \to \rho(\to\pi\pi)\ell^- \bar \nu_\ell$. However, the width of
the $K^*$-meson is sufficiently smaller than that of the $\rho$ resonance,
$\Gamma_{K^*} \simeq \Gamma_\rho / 4\simeq 50~\MeV$.
Moreover, from studies of the decay $\bar{B}\to \bar{K}^* J/\psi$ the S-wave
background below the $K^*$ resonance in $B$ decays is constrained to small
values, with the S-wave fraction $F_s \lesssim 7\%$ on-resonance
\cite{Aaij:2013cma}. The decay $\bar{B}_s \to K^{*+}(\to
K\pi)\ell^-\bar\nu_\ell$ thus provides a promising alternative channel
for a precise determination of $|V_{ub}|$ in the SM, 
as has already been advocated for in \cite{Meissner:2013pba}. For the same reason, 
it can also be used to constrain NP contributions in $b \to u$ transitions,
in particular, as we will show below, to exclude possible effects from right-handed
currents.

Another benefit of the decay $\bar{B}_s\to
K^{*+}\ell^-\bar\nu_\ell$ is the opportunity to 
combine it with the rare
$\bar{B}\to\bar{K}^*\ell^+\ell^-$ decay.
The secondary decay $K^*\to K\pi$ is identical in both decays, which
leads to a one-to-one correspondence between angular observables.
Hadronic form factors in both decays are related by the $SU(3)_f$ symmetry
of the strong interaction, and therefore hadronic uncertainties in \emph{ratios}
of angular observables from the two decays are expected to be under control.\footnote{%
The $B \to K^*\ell^+\ell^-$ decay amplitude also receives corrections from
nonfactorizable (i.e.\ not form-factor like) contributions involving hadronic 
operators in the $b \to s$ effective Hamiltonian. Semileptonic 
$b \to u$ transitions are free of such effects. A comparison of 
the two decays can thus also shed light on the size of 
nonfactorizable hadronic matrix elements and the validity of the underlying theoretical
framework. A detailed study along these lines is beyond the scope 
of the present work.}  

\clearpage 

Furthermore, these ratios of angular observables are sensitive not only to
$|V_{ub}|$, but also to bilinear combinations of the Wilson coefficients 
describing semileptonic $b \to u$ and radiative $b \to s$ transitions 
in the SM and beyond. In light of the present deviations
between LHCb measurements and the respective SM predictions 
for  a few angular observables in the $\bar{B}\to
\bar{K}^*$ channel (see e.g.\ \cite{Aaij:2013qta,Matias:2012xw}, and also \cite{Jager:2012uw}),
we will show how this can be exploited to obtain complementary information 
on the $b \to s$ Wilson coefficients.

The outline of the article is as follows. In \refsec{def}
we introduce the effective Hamiltonian for semileptonic $b\to u \ell \bar \nu_\ell$ 
transitions, including NP operators, and define the angular
observables for $\bar B_s \to K^*(\to K\pi)\ell \bar\nu_\ell$ transitions. 
In the following phenomenological section \ref{sec:phenoprop}
we identify SM null tests among the angular observables,
and derive expressions in a simplified scenario with only right-handed NP
contributions. We also define optimized observables and highlight the synergies
between the angular observables in $\bar B_s \to K^{*+}(\to K \pi)\ell\bar\nu_\ell$
and $\bar B \to K^*(\to K\pi) \ell^+\ell^-$.
In the numerical \refsec{numerics} we first perform a fit
of the Wilson coefficients for (V-A) and (V+A) currents 
to experimental data for $\bar{B} \to \pi^+ \ell^- \bar\nu_\ell$, 
$B^- \to \tau^- \bar \nu_\tau$ and $\bar{B} \to X_u \ell^- \bar\nu_\ell$ decays.
On the basis of this fit and theoretical estimates for the relevant form factors, 
we then provide numerical predictions for the angular observables and partially
integrated branching ratios for $\bar B_s \to K^{*+}(\to K\pi)\ell \bar\nu_\ell$
decays, before we conclude in \refsec{concl}.
The helicity basis for the $\bar{B}_s \to K^*$ form factors is defined
in \refapp{ff}, where we also infer the form factor parameters from light-cone
sum rule and lattice QCD results. The appendices \ref{app:hme} and
\ref{app:ang-dist-full} are dedicated to details on the determination of the
hadronic amplitudes and the angular observables 
of $\bar{B}_s\to K^{*+}\ell^-\bar\nu_\ell$ decays within and beyond the SM, 
respectively.

\section{Definitions}
\label{sec:def}

\subsection{Effective Hamiltonian for $b\to u \ell \bar\nu_\ell$}
\label{sec:pheno:eft}

We parametrize possible new physics contributions to $b\to u \ell \bar \nu_\ell$
transitions in a model-independent fashion in terms of a low-energy
effective Hamiltonian, which can be written in the form
\begin{equation}
  \label{eq:heff}
  \mathcal{H}^\mathrm{eff}_{b\to u}
    = -\frac{4 G_{\rm F} \Vubeff}{\sqrt{2}}\,  \sum_X \wilson{X} \, \op{X} + \text{ h.c.}\,.
\end{equation}
Here the most general set of dimension-6 operators  $\{\op{X}\}$
is given by
\begin{equation}
\label{eq:op-basis}
\begin{aligned}
  \op{V,i}
    & = \[\bar u \gamma^\mu P_i b\]\[\bar\ell \gamma_\mu P_L \nu_\ell\]\,,\\
  \op{S,i}
    & = \[\bar u P_i b\]\[\bar\ell P_L \nu_\ell\]\,,\\
  \op{T}
    & = \[\bar u \sigma^{\mu\nu} b\]\[\bar\ell \sigma_{\mu\nu} P_L\nu_\ell\]\,,\\
\end{aligned}
\end{equation}
where  $P_i\in\{P_L,P_R\}$ are chiral projectors,
and we have restricted ourselves to (massless) left-handed neutrinos and 
ignored the possibility of lepton-flavor violating couplings. (The generalization
to more exotic scenarios with light right-handed invisible neutral fermions is straight-forward,
see e.g.\ \cite{Dutta:2013qaa}.)
Since in the presence of NP the notion of $V_{ub}$ becomes ambiguous, we normalize 
the operators in \refeq{heff} to an
\emph{effective parameter} $\Vubeff$, which can be taken, for instance,
as the value of $V_{ub}$ that one obtains from a global CKM fit within the SM. 
If NP effects are small, one would then have $C_{V,L} \simeq 1$ (while in the SM
$C_{V,L}\equiv 1$ and $V_{ub} \equiv V_{ub}^{\rm eff}$, with all other Wilson
coefficients vanishing).
Comparing with reference \cite{Buras:2010pz},
where the modifications of left- and right-handed quark currents 
has been parametrized in terms of $\eps_{L,R}$ together with 
a new mixing-matrix $\tilde V$ for right-handed currents,
our conventions are related via
\begin{align}
    \left(\frac{       V_{ub} }{\Vubeff}\right) \eps_L & = \wilson{V,L} - 1\,, &
    \left(\frac{\tilde{V}_{ub}}{\Vubeff}\right) \eps_R & = \wilson{V,R}\,.
\end{align}

\subsection{Angular distribution in $\bar{B}_s\to K^* \ell \bar\nu_\ell$}
\label{sec:pheno:ang-dist}

The four-fold differential decay rate for $\bar{B}_s\to K^{*+}\ell^-\bar\nu_\ell$
is defined in terms of the dilepton invariant mass $q^2$, the polar angles 
$\theta_\ell$ and $\theta_{K^*}$ in the $\ell\nu$ and $K^*$ rest frames, respectively, 
and the azimuthal angle $\phi$  between
the primary and secondary decay planes,
\begin{equation}
  \label{eq:gamma}
  \frac{8\pi}{3} \frac{\dd^4\Gamma[\bar{B}_s\to K^*\ell^+\bar\nu_\ell]}{\dd q^2\,\dd\cos\theta_\ell\,\dd\cos\theta_{K^*}\,\dd\phi} = \hat{J}(q^2, \theta_\ell, \theta_{K^*}, \phi)\,.
\end{equation}
It can be expanded in a basis of trigonometric functions of the decay angles.
We define
\begin{align}\nonumber
  \Jhat(q^2, \theta_\ell, \theta_{K^*}, \phi)& = \Jhat_{1s} \sin^2\theta_{K^*} + \Jhat_{1c} \cos^2\theta_{K^*}\\
    &  + (\Jhat_{2s} \sin^2\theta_{K^*} + \Jhat_{2c} \cos^2\theta_{K^*}) \cos 2\theta_\ell \nonumber\\
    & + \Jhat_3 \sin^2\theta_{K^*} \sin^2\theta_\ell \cos 2\phi \nonumber\\
    & + \Jhat_4 \sin 2\theta_{K^*} \sin 2\theta_\ell \cos\phi \nonumber\\
    & + \Jhat_5 \sin 2\theta_{K^*} \sin\theta_\ell \cos\phi \nonumber\\
    & + (\Jhat_{6s} \sin^2\theta_{K^*}  + \Jhat_{6c} \cos^2\theta_{K^*}) \cos\theta_\ell \nonumber\\
    & + \Jhat_7 \sin 2\theta_{K^*} \sin\theta_\ell \sin\phi \nonumber\\
    & + \Jhat_8 \sin 2\theta_{K^*} \sin 2\theta_\ell \sin\phi \nonumber\\
    & + \Jhat_9 \sin^2\theta_{K^*} \sin^2\theta_\ell \sin 2\phi\,, 
  \label{eq:ang-dist}
\end{align}
with \emph{angular observables} $\Jhat_{i(a)} \equiv \Jhat_{i(a)}(q^2)$ for
$i=1,\dots,9$ and $a=s,c$. 
By construction, the functional
dependence of the angular distribution \refeq{ang-dist} on the angular observables
is identical to the one for $B\to V(\to P_1 P_2)\ell^+\ell^-$ decays in
\cite{Bobeth:2012vn}, to which we refer for further details.

Explicit expressions for the angular observables in terms of hadronic form factors 
and Wilson coefficients for $b \to u\ell\bar\nu_\ell$ in the general operator basis 
\eqref{eq:heff} are derived in
the appendices.

\section{Phenomenology}

\label{sec:phenoprop}

For the remainder
of this article we restrict our analysis to vector-like couplings; \ie, we
assume $\wilson{S,i} = \wilson{T} = \wilson{T5} = 0$ for simplicity.
This leaves us with only two operators for left- and right-handed $b\to u$ currents,
which we refer to as SM+SM'.
We emphasize that with future experimental data one can also test 
for scalar and tensor currents on the basis of the formulae provided
in \refapp{ang-dist-full}. 

\subsection{Null tests of the SM}

The twelve angular observables $\Jhat_i$ as introduced in \refeq{ang-dist} are not independent. Within
the SM, they can be expressed in terms of four real-valued quantities: $|N|^2$, and the
three form factors $F_{\perp,\para,0}$. This fact can be used to define a series of eight null
tests that hold within the SM:
\begin{align}
    4 \Jhat_{2c} \Jhat_{3} + \Jhat_5^2 - 4 \Jhat_4^2     & = 0\,,\cr
    8 \Jhat_{1s} \Jhat_{1c} - 3 \Jhat_5^2 - 12 \Jhat_4^2 & = 0\,,\cr
    \Jhat_{1c} \Jhat_{6s} - 2 \Jhat_4 \Jhat_5            & = 0\,,\cr
    16 \Jhat_{1s}^2 - 36 \Jhat_3^2 - 9 \Jhat_{6s}^2      & = 0\,,\cr
    \Jhat_{6c} = \Jhat_7 = \Jhat_8 = \Jhat_9             & = 0\,.
\end{align}
Deviations from these relations are immediate signs of physics  beyond the SM. This
is in contrast to exclusive $b\to s\ell^+\ell^-$ decays, where such relations are broken
by nonfactorizing long-distance contributions.

\subsection{Angular observables for SM+SM'}

In the SM+SM' scenario, we obtain  a
very simple structure of the angular observables, which can be expressed
in terms of hadronic form factors (defined in the transversity basis, see appendix~\ref{app:ff}),
and three independent combinations of Wilson coefficients,
\begin{equation}
\begin{aligned}
    \sigma_1^\pm & \equiv |\wilson{V,L} \pm \wilson{V,R}|^2\,,\\
     -2 \sigma_2 & \equiv (\wilson{V,L} - \wilson{V,R})(\wilson{V,L} + \wilson{V,R})^*\,,\\
\end{aligned}
\end{equation}
which depend on the absolute values $|C_{V,L}|$ and $|C_{V,R}|$ and the relative phase
of the two Wilson coefficients (the absolute phase is irrelevant in the angular
observables).
Notice that $\sigma_1^\pm$ is even under parity transformations ($ L \leftrightarrow R$),
while $\sigma_2$ is odd.
Neglecting the charged-lepton mass (which is valid as long as $m_\ell/\sqrt{q^2} \ll 1$),
we find
\begin{align}
    \Jhat_{1s} & = 3 \, \hat J_{2s} = 
     9 \, |N|^2 \, M_{B_s}^2 \Big[\sigma_1^+ \, |F_\perp|^2 + \sigma_1^- \, |F_\para|^2\Big]\,,\cr 
    \Jhat_{1c} & = - \hat J_{2c} = 12 \, |N|^2 \, \frac{M_{B_s}^4}{q^2} \, \sigma_1^- \, |F_{0}|^2\,,\cr 
    \Jhat_{3}  & = 6 \, |N|^2 \, M_{B_s}^2 \Big[\sigma_1^+ \, |F_\perp|^2 - \sigma_1^- \, |F_\para|^2\Big]\,,\cr 
    \Jhat_{4}  & = 6 \sqrt{2} \, |N|^2 \, \frac{M_{B_s}^3}{\sqrt{q^2}} \, \sigma_1^- \, F_\para \, F_0\,,
\end{align}
and 
\begin{align}
    \Jhat_{5}  & = 24 \, \sqrt{2} \,|N|^2\, \frac{M_{B_s}^3}{\sqrt{q^2}}\, \Re{\sigma_2}\, F_\perp \,F_0\,,\cr 
    \Jhat_{6s} & = 48 \,|N|^2 \,M_{B_s}^2\, \Re{\sigma_2}\, F_\perp\, F_\para\,,\cr
    \Jhat_{8}  & = 12\, \sqrt{2}\, |N|^2 \,\frac{M_{B_s}^3}{\sqrt{q^2}} \,\Im{\sigma_2} \,F_\perp\, F_0\,,\cr
    \Jhat_{9}  & = 24 \,|N|^2\, M_{B_s}^2\, \Im{\sigma_2}\, F_\perp\, F_\para\,,
\end{align}
together with $\Jhat_{6c} = \Jhat_{7} = 0$ (all relations valid in the SM+SM' scenario). 
Here, we introduce a normalization factor,
\begin{equation}
\label{eq:normalisation}
    |N|^2 \equiv \frac{G^2_\text{F} |V_{ub}^{\rm eff}|^2 q^2 \sqrt{\lambda}}{3 \cdot 2^{10} \pi^3 M_{B_s}^3}\,,
\end{equation}
and $\lambda \equiv \lambda(M_B^2, M_{K^*}^2, q^2)$ denotes the usual kinematic K\"all\'en function.
The normalization $|N|^2$ is chosen such that
\begin{align}
    \frac{\dd \Gamma}{\dd q^2}
    & = \sum_{\lambda=0,\perp,\para} |A^L_\lambda|^2\\
\nonumber
    & = |N|^2 M_{B_s}^2 \left[\sigma_1^+ |F_\perp| + \sigma_1^- \left(|F_\para|^2 + \frac{M_{B_s}^2}{q^2}|F_0|^2\right)\right]\,,
\end{align}
where the transversity amplitudes $A_\lambda^L$ are defined in \refapp{hme}.\\

Beside the decay rate, one can also define the leptonic forward-backward asymmetry $A_{\rm FB}$
via the weighted integral
\begin{equation}
A_{\rm FB} \equiv \frac{1}{\dd \Gamma / \dd q^2} \int_{-1}^{+1} \dd\cos\theta_\ell\,\sign(\cos\theta_\ell) \frac{\dd^2 \Gamma}{\dd q^2\,\dd \cos\theta_\ell}\,.
\end{equation}
In the SM+SM' scenario, one finds that $A_{\rm FB}$ takes the rather simple form
\begin{equation}
    A_{\rm FB} = \frac{2 \Re{\sigma_2} F_\perp F_\para}{\sigma_1^+ |F_\perp|^2 + \sigma_1^- \left(|F_\para|^2 + \frac{M_{B_s}^2}{q^2} |F_0|^2\right)}\,.
\end{equation}
Note, that the bilinear $\sigma_2$ is unconstrained by present experimental measurements
of semileptonic $b\to u$ transitions. Therefore a measurement of $A_{\rm FB}$ would
provide complementary information on the Wilson coefficients. In particular, the sign of
the forward-backward asymmetry resolves the present ambiguity between $\wilson{V,L}$ versus
$\wilson{V,R}$, see \refsec{numerics}.

Similarly, the fraction of longitudinal $K^*$ mesons is defined as
\begin{equation}
    F_L \equiv \frac{1}{\dd \Gamma / \dd q^2} \int_{-1}^{+1} \dd\!\cos\theta_{K^*}\,\omega_{F_L}(\cos\theta_{K^*}) \frac{\dd^2 \Gamma}{\dd q^2\,\dd\!\cos\theta_{K^*}}\,,
\end{equation}
where $\omega_{F_L}(z) = (5 z^2 - 1) / 2$. In the SM+SM' scenario this yields
\begin{equation}
    F_L = \frac{\sigma_1^- |F_0|^2}{\sigma_1^+ |F_\perp|^2 + \sigma_1^- \left(|F_\para|^2 + \frac{M_{B_s}^2}{q^2} |F_0|^2\right)}\,.
\end{equation}

\subsection{Optimized observables in SM+SM'}

It is now possible to construct particular combinations of 
angular observables where the hadronic form-factor dependencies
cancel (at least partially), and, as a consequence, these observables
are sensitive to the short-distance 
Wilson coefficients, only; or vice-versa.

We begin with observables where the form-factor dependencies cancel.
These can be defined in complete analogy to what has been discussed
in \cite{Bobeth:2012vn},
\begin{align}
    \hat H_T^{(1)} & = \frac{\sqrt{2} \Jhat_4}{\sqrt{-\Jhat_{2c} (2\Jhat_{2s} - \Jhat_3)}}\,,\cr 
    \hat H_T^{(2)} & = \frac{\Jhat_5}{\sqrt{-2\Jhat_{2c} (2\Jhat_{2s} + \Jhat_3)}}\,,\cr 
    \hat H_T^{(3)} & = \frac{\Jhat_{6s}}{2 \sqrt{(2\Jhat_{2c})^2 - (\Jhat_3)^2}}\,,\cr 
    \hat H_T^{(4)} & = \frac{2\Jhat_8}{\sqrt{-2\Jhat_{2c} (2\Jhat_{2s} + \Jhat_3)}}\,,\cr 
\label{eq:HT1bis5}
    \hat H_T^{(5)} & = \frac{-\Jhat_9}{\sqrt{(2\Jhat_{2c})^2 - (\Jhat_3)^2}}\,.
\end{align}

Within the SM+SM' scenario, the form factors dependencies cancel
\emph{exactly} at every point in the $q^2$ spectrum. However, 
for integrated angular observables one has to take into account the different 
kinematical prefactors, and a residual form-factor dependence will remain.\footnote{We
emphasize again that the cancellation of
form-factor dependencies holds for the whole $q^2$ spectrum, in contrast to $\bar{B} \to \bar K^*\ell^+\ell^-$
where it can be spoiled by contributions with intermediate photons dissociating into $\ell^+\ell^-$.}
In the SM+SM' scenario these optimized observables read 
\begin{equation}
\label{eq:HT-SM+SM'}
\begin{aligned}
    \hat H_T^{(1)}   & = 1\,,\\
    \hat H_T^{(2)} &= \hat H_T^{(3)}   = 2  \, \frac{\Re{\sigma_2}}{\sqrt{\sigma_1^+ \sigma_1^-}}\,,\\
    \hat H_T^{(4)} &= \hat H_T^{(5)}   = 2 \, \frac{\Im{\sigma_2}}{\sqrt{\sigma_1^+ \sigma_1^-}}\,.
\end{aligned}
 \end{equation}

We continue with the construction of observables that are \emph{only} sensitive to form-factor ratios.
Just as in $\bar{B}\to \bar{K}^*\ell^+\ell^-$, we find that the SM+SM' scenario solely allows us to 
extract one form factor ratio, namely $F_0 / F_\para$, in five different ratios of
angular observables,
\begin{multline}
    \frac{M_{B_s}}{\sqrt{q^2}} \frac{F_0(q^2)}{F_\para(q^2)} = \frac{\sqrt{2}\Jhat_5}{\Jhat_{6s}} = \frac{-\Jhat_{2c}}{\sqrt{2} J_4}\\
    = \frac{\Jhat_4}{2\Jhat_{2s} - \Jhat_3} = \sqrt{\frac{-\Jhat_{2c}}{2 \Jhat_{2s} - \Jhat_{3}}}= \frac{\sqrt{2} \Jhat_8}{-\Jhat_9}\,.
\end{multline}
Inconsistencies among these relations would indicate NP beyond the SM+SM' scenario.

\subsection{Synergies with $\bar{B} \to\bar{K}^*\ell^+\ell^-$}
\label{sec:pheno:syn}

The decay $\bar{B}\to\bar{K}^*(\to \bar{K}\pi)\ell^+\ell^-$ is induced by the
flavor-changing neutral current (FCNC) transition $b\to s\ell^+\ell^-$.
At low hadronic recoil, $q^2 \gtrsim 15~\GeV^2$, it is again dominated by 
four-fermion operators which can be extended to a SM+SM' scenario.
The structure of angular observables $J_n(q^2)$ in those decays 
is similar as for $\bar B_s \to \bar K^{*+}\ell\bar \nu_\ell$. 
The analogous combinations of Wilson coefficients which enter the $J_n(q^2)$
now read $\rho_1^\pm$ and $\rho_2$. 
(For the explicit definition and a detailed phenomenological discussion, 
we refer to \cite{Bobeth:2012vn}.)

With this we can define a number of useful ratios of angular observables
$J_n(q^2)$ in $\bar{B} \to \bar{K}^*\ell^+\ell^-$ and 
$\hat J_n(q^2)$ in $\bar{B}_s \to \bar K^{*+}\ell \bar\nu_\ell$,
\begin{equation}
    R_n(q^2) \equiv 
    \frac{J_n(q^2)
    }{\hat J_n(q^2)
    }\,,
\end{equation}
for $n=1c,2c,4,5,6s,8,9$, as well as
\begin{equation}
\begin{aligned}
    R_{1\pm}(q^2) & \equiv \frac{2J_{1s}(q^2) \pm 3 J_3(q^2)
    }{2\Jhat_{1s}(q^2) \pm 3 \Jhat_3(q^2) 
    }\,,\\
    R_{2\pm}(q^2) & \equiv \frac{2J_{2s}(q^2) \pm   J_3 (q^2)
    }{2\Jhat_{2s}(q^2) \pm   \Jhat_3(q^2)
    }\,.
\end{aligned}
\end{equation}
Within these ratios, the dependence on the hadronic form factors can be expected to cancel
up to corrections from the violation of the $SU(3)_f$ symmetry of strong interactions,
and from nonfactorizing hadronic matrix elements in exclusive $b\to s\ell^+\ell^-$
transitions.
In the limit where these corrections are neglected, we find 
\begin{equation}
    R_n \simeq \frac{\alpha_e^2}{8\pi^2} \, \frac{|V_{tb} V^*_{ts}|^2}{|V_{ub}|^2}
    \renewcommand{\arraystretch}{1.3}
    \begin{cases}
        \displaystyle \frac{\rho_1^+} {\sigma_1^+}      & \text{for }n = 1+, 2+\\
        \displaystyle \frac{\rho_1^-} {\sigma_1^-}      & \text{for }n = 1-, 1c, 2-, 2c\\
        \displaystyle \frac{\Re{\rho_2}}{\Re{\sigma_2}} & \text{for }n = 4, 5, 6s\\
        \displaystyle \frac{\Im{\rho_2}}{\Im{\sigma_2}} & \text{for }n = 8, 9\\
    \end{cases} \renewcommand{\arraystretch}{1.0}\,.
\end{equation}

Of particular interest are ratios that are proportional to the 
combination  $\rho_2 \propto \Re{\wilson{79}(q^2) \, \wilson[*]{10}}$,
where in the SM $\wilson{79}(q^2)$ is a linear combination of the Wilson coefficients
$\wilson[{\rm eff}]{7}$ and $\wilson[{\rm eff}]{9}(q^2)$ in $b \to s$ transitions
(see \cite{Bobeth:2012vn} for the explicit definitions).
Optimized observables in
$\bar{B}\to\bar{K}^*\ell^+\ell^-$ only allow to access 
the ratio $|\wilson[{\rm eff}]{9}/\wilson{10}|$, whereas 
the ratios $R_n$ are sensitive to  $\wilson[{\rm eff}]{9} \cdot \wilson{10}$.
Measuring the corresponding ratios $J_n/\hat J_n$ thus allows to directly 
access the $q^2$ dependence of $\wilson[{\rm eff}]{9}$ and to test the 
theoretical predictions which are based on an operator product expansion
in the heavy $b$-quark limit.

\section{Numerical Results}
\label{sec:numerics}

\begin{table}[t!!]
\renewcommand{\arraystretch}{1.3}
\resizebox{.49\textwidth}{!}{
\begin{tabular}{|c|c|c|c|}
    \hline
    Decay              & $q^2$ [GeV$^2$]  & Measurement                & Reference           \\
    \hline
    \multirow{4}{*}{$B^- \to \tau^-\bar\nu_\tau$}
    & --               & $(1.70 \pm 0.80 \pm 0.20) \cdot 10^{-4}$      & \cite{Aubert:2009wt}\\
    \cline{2-4}
    & --               & $(1.25 \pm 0.28 \pm 0.27) \cdot 10^{-4}$      & \cite{Kronenbitter:2015cha}  \\
    \cline{2-4}
    & --               & $(1.83^{+0.53}_{-0.59}\pm0.24) \cdot 10^{-4}$ & \cite{Lees:2012ju}  \\
    \cline{2-4}
    & --               & $(0.72^{+0.27}_{-0.25}\pm0.11) \cdot 10^{-4}$ & \cite{Adachi:2012mm}\\
    \hline
    \multirow{24}{*}{$\bar{B}^0 \to \pi^+ \mu^-\bar\nu_\tau$}
    & $[ 0,  2]$       & $(1.280 \pm 0.196) \cdot 10^{-5}$             & \multirow{6}{*}{\cite{delAmoSanchez:2010zd}}\\
    & $[ 2,  4]$       & $(1.192 \pm 0.135) \cdot 10^{-5}$             & \\
    & $[ 4,  6]$       & $(1.446 \pm 0.108) \cdot 10^{-5}$             & \\
    & $[ 6,  8]$       & $(1.437 \pm 0.105) \cdot 10^{-5}$             & \\
    & $[ 8, 10]$       & $(1.525 \pm 0.106) \cdot 10^{-5}$             & \\
    & $[10, 12]$       & $(1.490 \pm 0.111) \cdot 10^{-5}$             & \\
    \cline{2-4}
    & $[ 0,  2]$       & $(1.173 \pm 0.219) \cdot 10^{-5}$             & \multirow{6}{*}{\cite{Ha:2010rf}}\\
    & $[ 2,  4]$       & $(1.526 \pm 0.103) \cdot 10^{-5}$             & \\
    & $[ 4,  6]$       & $(1.213 \pm 0.105) \cdot 10^{-5}$             & \\
    & $[ 6,  8]$       & $(1.465 \pm 0.102) \cdot 10^{-5}$             & \\
    & $[ 8, 10]$       & $(1.473 \pm 0.108) \cdot 10^{-5}$             & \\
    & $[10, 12]$       & $(1.404 \pm 0.124) \cdot 10^{-5}$             & \\
    \cline{2-4}
    & $[ 0,  2]$       & $(1.225 \pm 0.182) \cdot 10^{-5}$             & \multirow{6}{*}{\cite{Lees:2012vv}}\\
    & $[ 2,  4]$       & $(1.277 \pm 0.128) \cdot 10^{-5}$             & \\
    & $[ 4,  6]$       & $(1.274 \pm 0.109) \cdot 10^{-5}$             & \\
    & $[ 6,  8]$       & $(1.498 \pm 0.103) \cdot 10^{-5}$             & \\
    & $[ 8, 10]$       & $(1.405 \pm 0.115) \cdot 10^{-5}$             & \\
    & $[10, 12]$       & $(1.617 \pm 0.104) \cdot 10^{-5}$             & \\
    \cline{2-4}
    & $[ 0,  2]$       & $(1.95 \pm 0.32) \cdot 10^{-5}$               & \multirow{6}{*}{\cite{Sibidanov:2013rkk}}\\
    & $[ 2,  4]$       & $(1.06 \pm 0.27) \cdot 10^{-5}$               & \\
    & $[ 4,  6]$       & $(1.51 \pm 0.28) \cdot 10^{-5}$               & \\
    & $[ 6,  8]$       & $(0.97 \pm 0.23) \cdot 10^{-5}$               & \\
    & $[ 8, 10]$       & $(0.78 \pm 0.22) \cdot 10^{-5}$               & \\
    & $[10, 12]$       & $(1.59 \pm 0.28) \cdot 10^{-5}$               & \\
    \hline
\end{tabular}
}
\caption{
    Summary of the experimental likelihoods for branching fractions of the
    exclusive $b\to u$ transitions.  We assume no correlation among the
    $B^-\to\tau^-\bar\nu_\tau$ data, and use the correlation matrices as given
    in \cite[tables XI and XII]{delAmoSanchez:2010zd}, \cite[table III and
    IV]{Ha:2010rf}, \cite[tables XXIX and XXXII]{Lees:2012vv} and \cite[table
    XVII]{Sibidanov:2013rkk} for the respective data on $\bar{B}^0\to
    \pi^+\mu^-\bar\nu_\mu$ decays.
    \label{tab:wc-constraints}
}
\end{table}

In this section we derive numerical results for the angular
observables $\Jhat_n$ as introduced in \refsec{pheno:ang-dist}. Our analysis
is carried out within a Bayesian framework, for which we use and extend EOS \cite{EOS}
for all numerical evaluations. As prerequisites to our numerical study of the angular observables,
information on the $\bar{B}_s \to K^*$ form factors,
and constraints on the $b\to u$ Wilson coefficients are needed. These will be expressed
through \emph{a-posteriori} probability density functions (PDFs) labelled
$P(\vec\theta_\text{FF} | \text{theory})$ and $P(\vec\theta_{\Delta B} | \text{exp. data})$,
respectively. We refer to \refapp{ff} for the precise definition of $P(\vec\theta_\text{FF} | \text{theory})$.\\

\subsection{Determination of $C_{V,L}$ and $C_{V,R}$}

For the following numerical analysis we consider 
experimental data on the branching ratios for leptonic $B^-\to \tau^-\bar\nu_\tau$ and
semileptonic $\bar{B^0}\to \pi^+\mu^-\bar\nu_\mu$ decays 
as summarized in table~\ref{tab:wc-constraints},
together with the averaged value for $|V_{ub}|$ from the inclusive determination,
\begin{equation}
\label{eq:inclusive}
\begin{aligned}
    |V_{ub}^\text{incl.}|
        & = (4.41 \pm 0.21) \cdot 10^{-3}\quad\text{\cite{Kowalewski:2014PDG}}\,.
\end{aligned}
\end{equation}
Within the SM+SM' scenario, the additional right-handed operator
contributes differently to the individual decay rates, corresponding to 
(see \eg \cite{Buras:2010pz})
\begin{equation}
\begin{aligned} 
|V_{ub}^{B \to \tau \nu}|^2 & \to  |\Vubeff|^2 \left|\wilson{V,L} - \wilson{V,R}\right|^2 \,,
\cr 
|V_{ub}^{B \to \pi \ell \nu}|^2 & \to  |\Vubeff|^2 \left|\wilson{V,L} + \wilson{V,R}\right|^2 \,,
\cr 
|V_{ub}^\text{incl.}|^2 & \to  |\Vubeff|^2 \left(|\wilson{V,L}|^2 + |\wilson{V,R}|^2\right) \,.
\end{aligned}
\end{equation}
In order to illustrate the NP reach of our analysis, we fix 
the auxiliary parameter $\Vubeff$
to a value that lies between the exclusive and inclusive 
determinations of $|V_{ub}|$ within the SM, 
$$|\Vubeff| \equiv 3.99\cdot 10^{-3} \,.$$
With this we can constrain the absolute values and the relative phases
of the Wilson coefficients $\wilson{V,L}$ and $\wilson{V,R}$, where
the SM-like solution would correspond to $|\wilson{V,L}|\sim 1$ and $\wilson{V,R}\sim 0$.

We construct a likelihood $P(\text{data} | \vec\theta_{\Delta B}, M)$ from
(multi)normal distributions as indicated in \reftab{wc-constraints} and
\refeq{inclusive}.  Note that we assume that the results for the $B^-\to
\tau^-\bar\nu_\tau$ branching ratios \cite{Aubert:2009wt} and
\cite{Lees:2012ju} are uncorrelated, since the underlying sets of events use
different tagging methods for the selection process.
The same assumption applies to the results of
\cite{Adachi:2012mm} and \cite{Hara:2010dk}. 
At this time, we only use theoretical
input from light-cone sum rules for the $B \to \pi$ transition form factors,
and therefore restrict ourselves to the kinematic range $q^2 \leq 12~\GeV^2$.
For a consistent inclusion of 
lattice results on the $B \to \pi$ form factor in the high-$q^2$ 
region (see \eg \cite{Dalgic:2006dt,Bailey:2008wp,AlHaydari:2009zr}, but also note added below),
we presently do not have access to the necessary correlation information required
for our statistical procedure.

Within our analysis, we address the theoretical uncertainties using nuisance
parameters for the hadronic matrix elements. These are the $B$-meson decay
constant $f_{B^-}$, and the parameters of the the $B\to \pi$ vector form factor
$f^{B\pi}_+(q^2)$: its normalization $f^{B\pi}_+(0)$, as well as two shape
parameters $b^{B\pi}_{1,2}$; see \cite{Imsong:2014oqa} for their definition.
For the $B$-meson decay constant we use a Gaussian prior with central value and
minimal $68\%$ probability interval $f_{B^-} = (210\pm 11)~\MeV$, as obtained
from a recent 2-point QCD sum rule at NNLO accuracy \cite{Gelhausen:2013wia}.
As prior for the form factor parameters we use the a-posteriori distribution
obtained from a recent Bayesian analysis of the LCSR prediction at NLO accuracy
\cite{Imsong:2014oqa}.

In order to assess the physical implications of possible deviations from the SM expectations,
we compare the fit results for three different scenarios. In all cases we assume 
$C_{V,L}$ to be real-valued (i.e.\ a possible NP phase in the left-handed 
$b \to u$ transition should be associated to $V_{ub}^{\rm eff}$).
As already mentioned, the fit to the considered data is only sensitive to
the relative phase between the Wilson coefficients $C_{V,L}$ and $C_{V,R}$,
and consequently we will always encounter an irreducible degeneracy related 
to $C_{V,L/R} \to - C_{V,L/R}$.
\begin{enumerate}
\item First, we consider the scenario ``left'' that features only the left-handed
current. In this case the number of parameters is five,
$
    \vec\theta^\text{left}_{\Delta B}
    = \left( \wilson{V,L}, f^{B\pi}_+(0), b^{B\pi}_1, b^{B\pi}_2, f_{B^-} \right)$.
\item
Next, we consider the scenario ``real'', in which $\wilson{V,R}$ is present
and real-valued. The set of $\Delta B$ parameters then reads
$
    \vec\theta^\text{real}_{\Delta B}
    = \left( \wilson{V,L}, \operatorname{Re}\wilson{V,R}, f^{B\pi}_+(0), b^{B\pi}_1, b^{B\pi}_2, f_{B^-} \right)$.
\item Last but not least, we also consider the scenario ``comp'', which includes
a complex-valued $\wilson{V,R}$, with 
the full seven parameters,
 $   \vec\theta^\text{comp}_{\Delta B} = \big( \wilson{V,L}, \operatorname{Re}\wilson{V,R}, \operatorname{Im}\wilson{V,R},
    f^{B\pi}_+(0), b^{B\pi}_1, b^{B\pi}_2, f_{B^-} \big)$.
\end{enumerate}

For all scenarios ($M = \text{left}, \text{real}, \text{comp}$), we obtain the
\emph{a-posteriori} PDF as usual via Bayes' theorem,
\begin{equation}
    P(\vec\theta_{\Delta B} | \text{data}, M) = \frac{P(\text{data} | \vec\theta_{\Delta B}, M) P_0(\vec\theta_{\Delta B}, M)}{P(\text{data}, M)}\,,
\end{equation}
where
\begin{equation}
    P(\text{data}, M) \equiv \int \dd\vec\theta_{\Delta B}\, P(\text{data} | \vec\theta_{\Delta B}, M) P_0(\vec\theta_{\Delta B}, M)
\end{equation}
is the evidence for the scenario $M$. The likelihood $P(\text{data} |
\vec\theta_{\Delta B}, M)$ has already been introduced earlier.
In all three scenarios, we use for the priors of the Wilson coefficients
uncorrelated, uniform distributions with the support $-2 \leq \wilson{i} \leq
+2$. For model comparisons, we normalize the model priors for the various fits
scenarios. The corresponding relations read
\begin{equation}
    P_0(\text{comp}) : P_0(\text{real}) : P_0(\text{left}) = 1 : 4 : 16
\end{equation}

\begin{table}
    \renewcommand{\arraystretch}{1.3}
    \begin{tabular}{|c|c|c|c|c|c|}
        \hline
                           & \multicolumn{3}{c|}{Significance [$\sigma$]}  & &                   \\
        Quantity           & ``left'' & ``real''& ``comp''        & d.o.f. & Reference           \\
        \hline
        $f_+ (\dagger)$    & $ 3.11$  & $ 2.36$ & $ 2.36$         & 3      & \cite{Imsong:2014oqa}\\
        \hline
        \multirow{4}{*}{$B^- \to \tau^-\bar\nu_\tau$}
                           & $+0.57$  & $+0.39$ & $+0.39$         & 1      & \cite{Aubert:2009wt}\\
        \cline{2-4}
                           & $+0.64$  & $+0.34$ & $+0.34$         & 1      & \cite{Kronenbitter:2015cha}  \\
        \cline{2-4}
                           & $+0.99$  & $+0.75$ & $+0.75$         & 1      & \cite{Lees:2012ju}  \\
        \cline{2-4}
                           & $-1.84$  & $-2.35$ & $-2.35$         & 1      & \cite{Adachi:2012mm}\\
        \hline
        \multirow{4}{*}{$\bar{B}^0 \to \pi^+ \mu^-\bar\nu_\tau$}
                           & $ 0.85$  & $ 1.08$ & $ 1.08$         & 6      & \cite{delAmoSanchez:2010zd}\\
        \cline{2-4}
                           & $ 0.87$  & $ 0.98$ & $ 0.98$         & 6      & \cite{Ha:2010rf}\\
        \cline{2-4}
                           & $ 1.70$  & $ 1.97$ & $ 1.97$         & 6      & \cite{Lees:2012vv}\\
        \cline{2-4}
                           & $ 2.53$  & $ 2.46$ & $ 2.46$         & 6      & \cite{Sibidanov:2013rkk}\\
        \hline
        $\bar{B}\to X_u\ell^-\bar\nu_\ell$
                           & $+1.67$  & $+1.45$ & $+1.45$         & 1      & \cite{Kowalewski:2014PDG}\\
        \hline
    \end{tabular}
    \renewcommand{\arraystretch}{1.3}
    \caption{Significances of the measurements at the best-fit point closest to
        the SM point for all three fit scenarios. Notice that the pull for the
        LCSR calculation of the $B\to\pi$ vector form factor $f_+$, marked by a
        $\dagger$, does not enter the goodness-of-fit calculation.
    }
\label{tab:pull-values}
\end{table}

\subsubsection{Scenario ``Left''}

Our findings for the scenario ``left'' can be summarized as follows.
We find two degenerate best-fit points corresponding to $|\wilson{V,L}| \simeq 1$.
The best-fit point (with positive $C_{V,L}$) reads
\begin{equation}
    \vec\theta_{\Delta B}^\text{left,$*$} =\\
    (1.016, 0.232, -3.163, +0.425, 0.206)\,.
\end{equation}
We find at this point $\chi^2_\text{left} = 18.54$, for $28$ degrees of freedom
(from $29$ measurements reduced by $1$ fit parameter). As a consequence, this
represents an excellent fit with a p-value of $91\%$. The significances of the
individual experimental inputs are collected in \reftab{pull-values}.
The one-dimensional marginalized posterior is approximately Gaussian, and yields
\begin{equation}
    |\wilson{V,L}| = 1.02 \pm 0.05\quad \text{at $68\%$ probability}\,.
\end{equation}
Equivalently, this result can be expressed as $|V_{ub}| = (4.07 \pm 0.20)\cdot
10^{-3}$ at $68\%$ probability.

\subsubsection{Scenario ``Real''}

\begin{figure*}[p!t]
    \includegraphics[width=.49\textwidth]{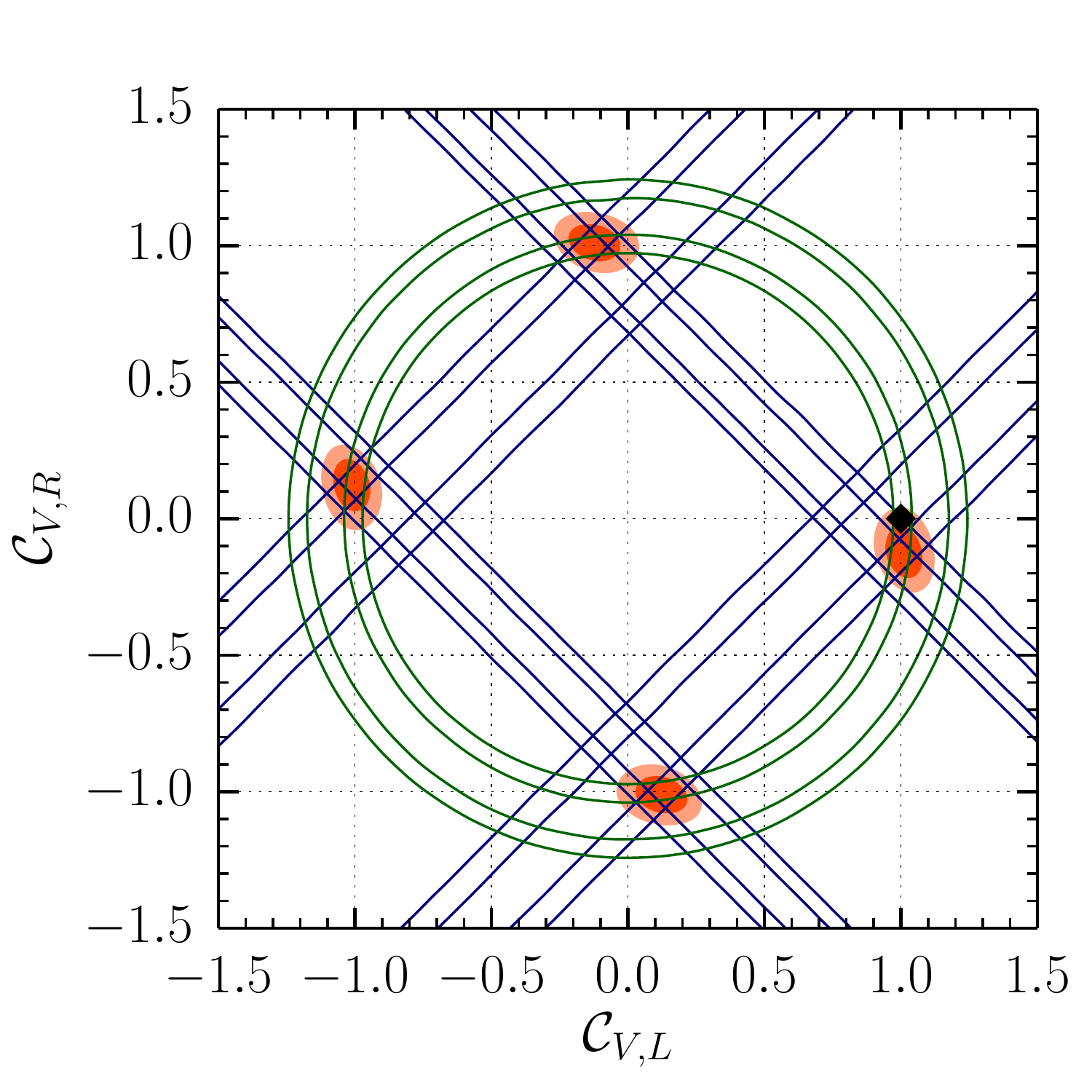}
    \includegraphics[width=.49\textwidth]{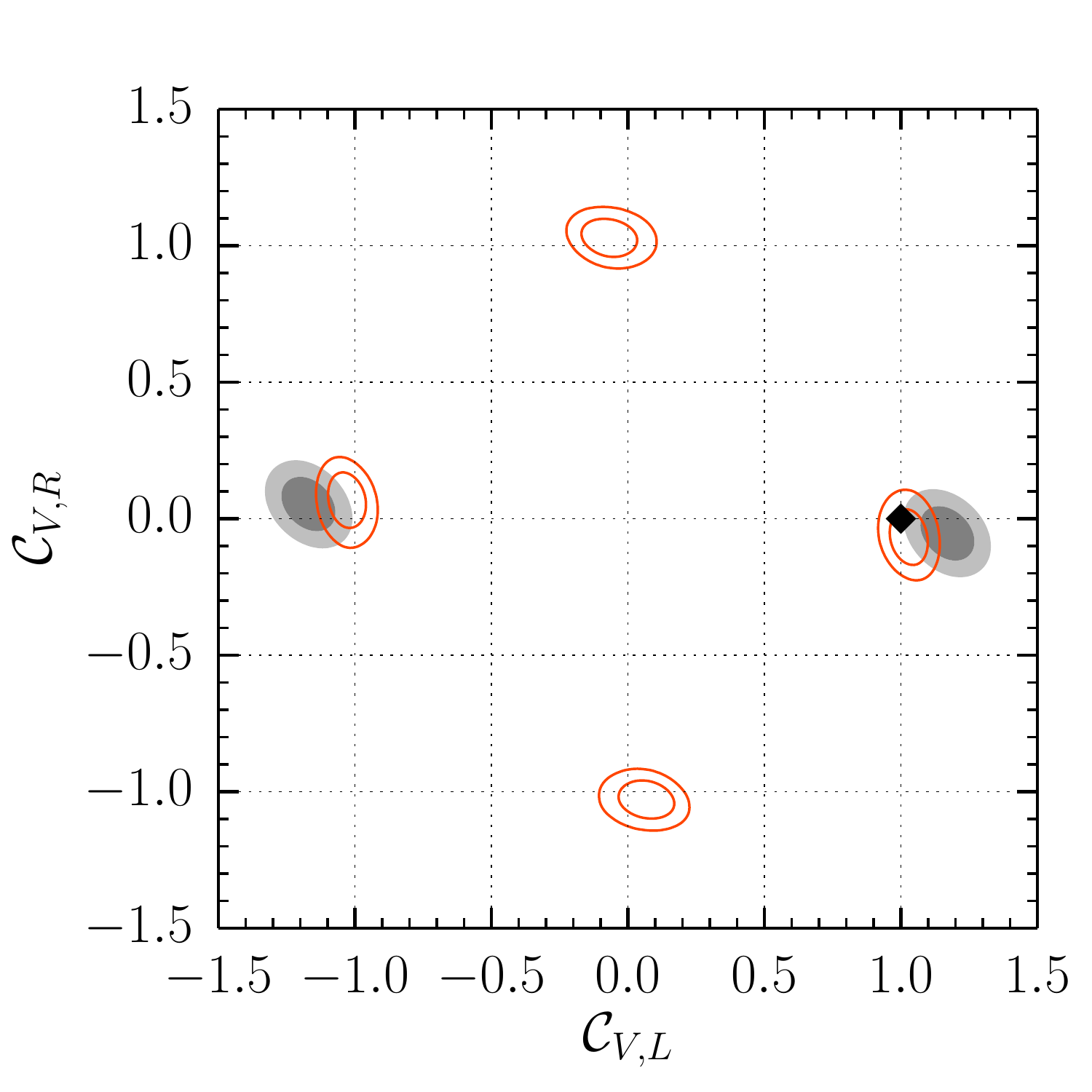}
    \caption{
        (left) Contours of the $68\%$ (dark orange area) and $95\%$ (orange
        area) probability regions for the Wilson coefficients $\wilson{V,L}$
        and $\wilson{V,R}$ as obtained from our fit. See the text for details.
        Overlaid are the $68\%$ and $95\%$ contour lines for $\bar{B}^0\to \pi^+\ell^-\bar\nu_\ell$
        (blue solid lines, negative slope), $B^- \to \ell^-\bar\nu_\ell$ (blue solid lines, positive slope)
        and inclusive $\bar{B}\to X_u \ell^-\bar\nu_\ell$ (green solid rings).
        The black diamond marks the SM point.
        (right) Contours of the $68\%$ and $95\%$ probability regions for the
        Wilson coefficients (solid orange lines) overlaying the $68\%$ (dark
        gray area) and $95\%$ (light gray area) probability regions as obtained
        from a hypothetical measurement of $A_{\rm FB} = A_{\rm FB}^\text{SM} \pm 10\%$.
    }
\label{fig:wc-constraints}
\end{figure*}
\begin{figure*}[p!!bth]
    \includegraphics[width=.49\textwidth]{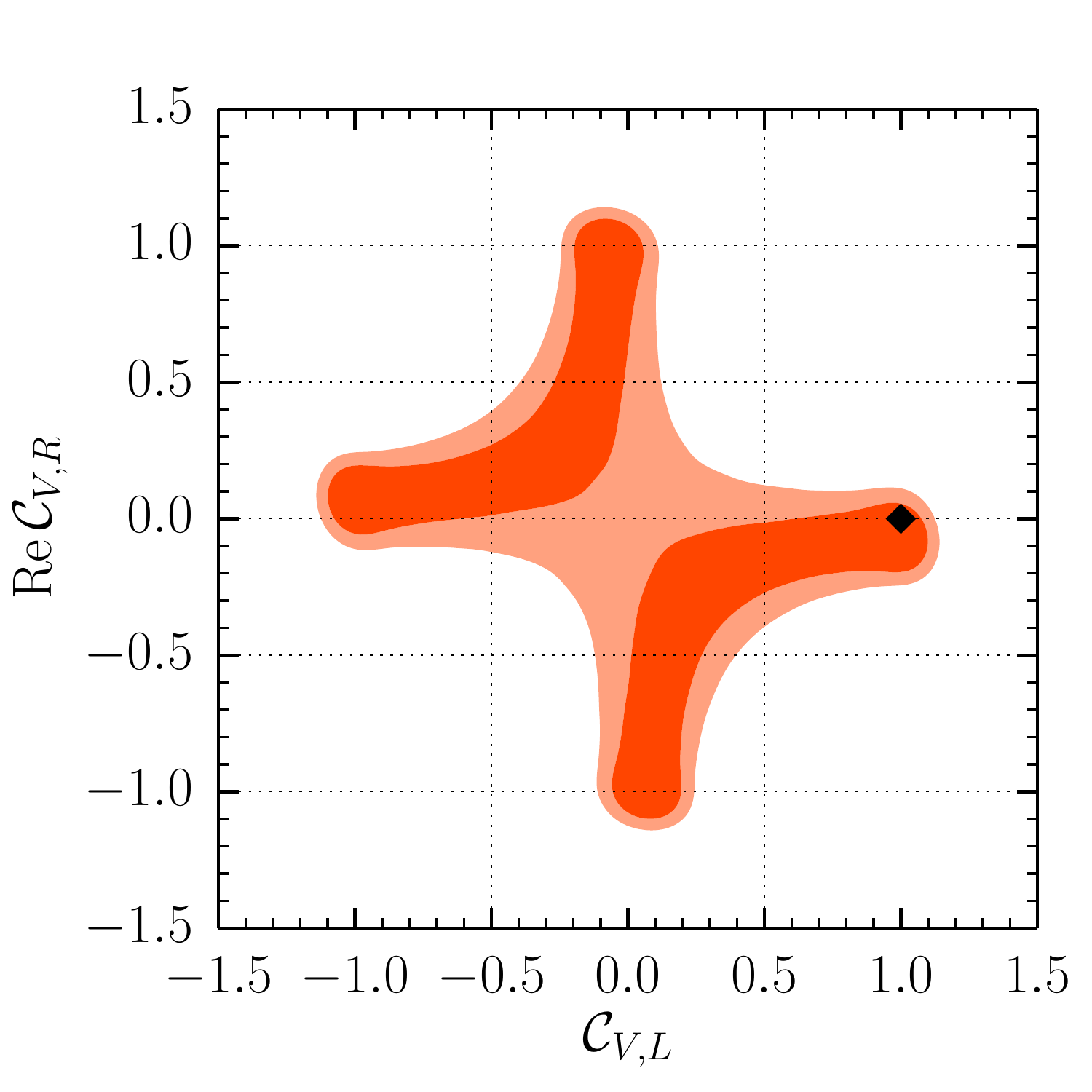}
    \includegraphics[width=.49\textwidth]{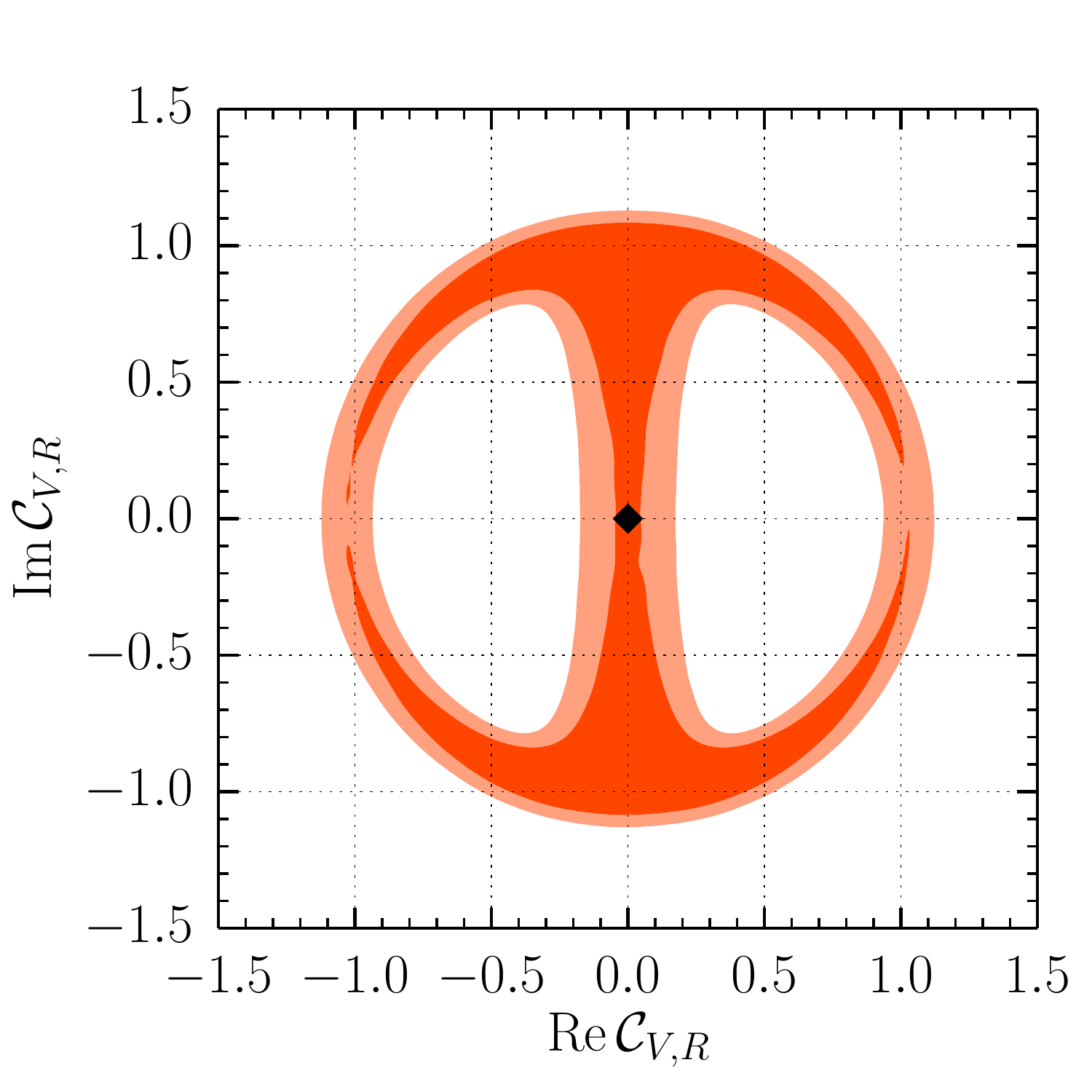}
    \caption{
        Contours of the $68\%$ (dark orange area) and $95\%$ (orange area)
        probability regions for the Wilson coefficients $\wilson{V,L}$ and
        $\wilson{V,R}$ as obtained from our fit in scenario ``comp''.  See the
        text for details. The black diamond marks the SM point.
    }
\label{fig:wc-constraints-complex}
\end{figure*}
For the scenario ``real'', we find a four-fold ambiguity in the data; see
\reffig{wc-constraints} for an illustration. All local modes are degenerate.
We calculate the goodness of fit in the local mode closest to the SM,
\begin{multline}
    \vec\theta_{\Delta B}^\text{real,$*$} =\\
    (1.025, -0.079, 0.251, -2.884, +0.196, 0.200)\,,
\end{multline}
and obtain $\chi^2_\text{real} = 20.47$. This fit's p-value of $81\%$ is very
good. However, note that the $\chi^2$ value has increased in comparison to the
previous scenario. This result warrants a comment. The additional degree
of freedom in form of $\wilson{V,R}$ allows the fit to move the form factor
parameters $f^+_{B\pi}$, $b_1$ and $b_2$ closer to the central values of the
prior. This shift occurs at the expense of increasing 
the significances of the experimental data, 
while simultaneously reducing the significance of the nuisance parameters.
For completeness, we also list these significances for all scenarios in
\reftab{pull-values}. The one-dimensional marginalized posterior distributions
for this scenario are approximately Gaussian and symmetric under exchange
$\wilson{V,L} \leftrightarrow \operatorname{Re}\wilson{V,R}$. We find
(at 68\%\ propabality)
\begin{align}
    |\wilson{V,L}| = 1.02 \pm 0.05\quad & \text{and}  \quad |\operatorname{Re} C_{V,R}| \leq  0.10
     \,, 
     \label{eq:real1}\\
&\mbox{or} \cr     
     |\operatorname{Re}\,\wilson{V,R}|
     = 1.02 \pm 0.05\quad &\text{and} \quad |\wilson{V,L}| \leq 0.10  \,.
     \label{eq:real2}
\end{align}

\subsubsection{Scenario ``Comp''}

We repeat the fit in scenario ``comp''. As a consequence of the additional
degree of freedom, the four solutions from the previous scenario now become
connected. This is illustrated in \reffig{wc-constraints-complex}.  We
calculate the goodness of fit in the local mode closest to the SM, which now
reads:
\begin{multline}
    \vec\theta_{\Delta B}^\text{comp,$*$} =\\
    (1.025, -0.080, 0.000, 0.251, -2.885, +0.196, 0.200)\,.
\end{multline}
The individual significances are listed in \reftab{pull-values}, and amount to
a total $\chi^2 = 20.48$. For the increase of $\chi^2$ with respect to the
``left'' scenario, see our earlier comment.
With $26$ degrees of freedom the p-value is $77\%$, which is still very good.
It is not sensible to provide the $68\%$ probability interval of the one-dimensional
marginalized posterior, since the solutions are strongly connected. We show
the contours of the probability regions at $68\%$ and $95\%$ probability
in \reffig{wc-constraints-complex}.


\subsubsection{Comparison}

We proceed with a comparison of the various fit scenarios by means of the
posterior odds. The latter can be calculated as
\begin{equation}
    \frac{P(M_1|\text{data})}{P(M_2|\text{data})}
    = \frac{P(\text{data}|M_1)}{P(\text{data}|M_2)}
      \frac{P_0(M_1)}{P_0(M_2)}
\end{equation}
We find
\begin{equation}
    \frac{P(\text{``left''}|\text{data})}{P(\text{``real''}|\text{data})}
    = 27.8 \, : \, 1\,,
\end{equation}
and
\begin{equation}
    \frac{P(\text{``real''}|\text{data})}{P(\text{``comp''}|\text{data})}
    = 3.62 \, : \, 1\,.
\end{equation}
Using Jeffrey's scale for the interpretation of the posterior odds
\cite{Jeffreys:1961}, we find that the data favour the interpretation with
purely left-handed $b\to u$ currents over the other scenarios \emph{very
strongly}. Moreover, the scenario ``real'' is \emph{substantially favored} over
the scenario ``comp''.

This means that -- despite the observed tensions between the different SM determinations
of $|V_{ub}|$ -- a NP scenario with right-handed currents does not lead to a more
efficient description of the experimental data. We emphasize again that the
statistical treatment of the 
theoretical uncertainties on the hadronic input parameters, which are still
relatively large at present, has been crucial for this argument. 
On the other hand, the experimental data on the inclusive and exclusive
decay rates alone also cannot \emph{exclude} large right-handed currents.

\subsection{Predictions for Angular Observables $\Jhat_n$}

We can now proceed to produce predictive distributions for the angular
observables $\Jhat_n$ in $\bar B_s \to K^{*+}(\to K\pi) \ell \bar \nu_\ell$,
for which we have two main applications in mind.

\subsubsection{SM Scenario}

First, we assume the SM case; i.e., we go back to  $V_{ub}^{\rm eff} \to V_{ub}$ 
with $\wilson{V,L} \equiv 1$ and $\wilson{i}
\equiv 0$. In this case, only the \emph{a-posteriori} PDF on the $\bar{B}_s\to K^*$
form factors is needed.  We obtain the joint posterior-predictive distribution
for the angular observables by means of
\begin{equation}
    P(\vec{\Jhat}) = \int \dd\vec\theta_\text{FF}\, P(\vec\theta_\text{FF} | \text{theory}) \,
    \delta(\vec{\Jhat} - \Jhat(\vec\theta_\text{FF}))\,.
\end{equation}
In practice, the above is carried out by calculating the $\Jhat_n$ for a set of
samples drawn from the \emph{a-posteriori} PDF. In our analysis $10^6$ samples
are used.
Our results for the angular observables, normalized to the decay width, are
compiled in \reftab{jhat-sm}. We single out the branching ratio, which appears
to be the most immediate candidate for upcoming measurement. We
present our results in units of $|V_{ub}|^{-2}$, which is convenient
to extract $|V_{ub}|$ from future data.  Our
results read
\begin{align}
    \int_{1\,\GeV^2}^{6\,\GeV^2} \dd q^2\, \frac{\dd \mathcal{B}}{\dd q^2}
    & = (5.08^{+0.95}_{-0.64})\, |V_{ub}|^{-2}\,,\cr 
    \int_{14.18\,\GeV^2}^{q^2_\text{max}} \dd q^2\, \frac{\dd \mathcal{B}}{\dd q^2}
    & = (8.50^{+0.29}_{-0.32})\, |V_{ub}|^{-2}\,,\cr 
    \int_{q^2_\text{min}}^{q^2_\text{max}} \dd q^2\, \frac{\dd \mathcal{B}}{\dd q^2}
    & = (27.25^{+2.15}_{-1.93})\, |V_{ub}|^{-2}\,.
\end{align}
In the above, $q^2_\text{min} = 0.02$, and $q^2_\text{max} = (M_{B_s} - M_{K^*})^2$.\\

\subsubsection{SM+SM' Scenario}

Second, we consider the interesting prospect of NP effects entering the $b\to u$
transitions, which, according to the discussion in the previous subsection,
cannot yet be ruled out.  Based upon our model comparison, we choose to give predictions
for the scenario ``real'' only.  In order to investigate the NP effects on the
angular observables in $\bar{B}_s\to K^*\ell\bar\nu_\ell$, we compute the joint
predictive distribution that arises from both posteriors $P(\vec\theta_{\Delta
B}|\text{data})$ and $P(\vec\theta_\text{FF}|\text{theory})$.
Our findings are listed in \reftab{jhat-np} for our three nominal choices of
$q^2$ bins. In addition, we find for the partially integrated branching ratios
in the scenario ``real''
\begin{align}
    \int_{1\,\GeV^2}^{6\,\GeV^2} \dd q^2\, \frac{\dd \mathcal{B}}{\dd q^2}
    & = (9.5 \pm 1.9) \cdot 10^{-5}\,,\cr 
    \int_{14.18\,\GeV^2}^{q^2_\text{max}} \dd q^2\, \frac{\dd \mathcal{B}}{\dd q^2}
    & = (1.55 \pm 0.19) \cdot 10^{-4}\,,\cr
    \int_{0.02\,\GeV^2}^{q^2_\text{max}} \dd q^2\, \frac{\dd \mathcal{B}}{\dd q^2}
    & = (4.92 \pm 0.69) \cdot 10^{-4}\,.
\end{align}

We also consider suitable ratios of partial decay widths in $\bar{B}_s \to K^{*+}\mu^-\bar\nu_\mu$
over either the $\bar{B}^0\to\pi^+\mu^-\bar\nu_\mu$ or $B^-\to \tau^-\bar\nu_\tau$ widths.
We define three such ratios,
\begin{align}
    \tilde R_{0}   & \equiv \frac{\int_{q^2_\text{min}}^{q^2_\text{max}} \dd q^2\, |A_{0    }^L|^2}{\Gamma(B^-\to\tau^-\bar\nu_\tau)} = \frac{3\Jhat_{1c} - \Jhat_{2c}}{3\Gamma(B^-\to\tau^-\bar\nu_\tau)}\,,\cr
    \tilde R_\para & \equiv \frac{\int_{q^2_\text{min}}^{q^2_\text{max}} \dd q^2\, |A_{\para}^L|^2}{\Gamma(B^-\to\tau^-\bar\nu_\tau)} = \frac{8\Jhat_{1s} - 12\Jhat_3}{9\Gamma(B^-\to\tau^-\bar\nu_\tau)}\,,\cr
    \tilde R_\perp & \equiv \frac{\int_{q^2_\text{min}}^{q^2_\text{max}} \dd q^2\, |A_{\perp}^L|^2}{\langle \Gamma(\bar{B}^0\to \pi^+ \mu^-\bar\nu_\mu)\rangle}\cr
                   & = \frac{8\Jhat_{1s} + 12\Jhat_3}{9\langle \Gamma(\bar{B}^0\to\pi^+\mu^-\bar\nu_\mu)\rangle}\,,
\end{align}
where, as already explained above, we only use the LCSR-accessible part of the $\bar{B}^0\to \pi^+\mu^-\bar\nu_\mu$
phase space,
\begin{multline}
    \langle \Gamma(\bar{B}^0\to\pi^+\mu^-\bar\nu_\mu)\rangle\\
    = \int_{q^2_\text{min}}^{12\,\GeV^2} \dd q^2\, \frac{\dd \Gamma(\bar{B}^0\to\pi^+\mu^-\bar\nu_\mu)}{\dd q^2}\,.
\end{multline}
The ratios $\tilde R_{0,\para,\perp}$ are independent of NP effects in this scenario.
We find numerically,
\begin{align}
    \tilde R_{0}     & = 2.00^{+0.39}_{-0.32}\,,\cr
    \tilde R_{\para} & = 1.36^{+0.17}_{-0.14}\,,\cr
    \tilde R_{\perp} & = 0.79^{+0.14}_{-0.10}\,,
\end{align}
where the uncertainties are purely due to the imprecise theoretical knowledge
of the $\bar{B}_s\to K^*$ form factors, the $\bar{B}\to \pi$ form factors, and
the $B$-meson decay constant. Here, correlation information among the various
hadronic matrix elements would help in reducing these uncertainties.

\begin{table}[t]
    \renewcommand{\arraystretch}{1.3}
    \begin{tabular}{%
        | @{\quad} c @{\quad}%
        | S[table-format=+3.3] @{\,${}^{+}_{-}$\,} S[table-format=1.7] %
        | S[table-format=+3.3] @{\,${}^{+}_{-}$\,} S[table-format=1.7] %
        | S[table-format=+3.3] @{\,${}^{+}_{-}$\,} S[table-format=1.7] %
        |
    }
        \hline
            & \multicolumn{6}{c|}{$\langle \Jhat_n\rangle / \langle \Gamma\rangle$}\\
        $n$
            & \multicolumn{2}{c|}{(a)}
            & \multicolumn{2}{c|}{(b)}
            & \multicolumn{2}{c|}{(c)}
        \\
        \hline
        $1s$
            & 0.144 & ${}^{0.020}_{0.020}$
            & 0.368 & ${}^{0.008}_{0.006}$
            & 0.283 & ${}^{0.018}_{0.020}$
        \\
        \hline
        $1c$
            & 0.558 & ${}^{0.027}_{0.027}$
            & 0.260 & ${}^{0.008}_{0.010}$
            & 0.373 & ${}^{0.026}_{0.024}$
        \\
        \hline
        $2s$
            & 0.048 & ${}^{0.007}_{0.007}$
            & 0.123 & ${}^{0.003}_{0.002}$
            & 0.094 & ${}^{0.006}_{0.007}$
        \\
        \hline
        $2c$
            &-0.558 & ${}^{0.027}_{0.027}$
            &-0.260 & ${}^{0.010}_{0.008}$
            &-0.373 & ${}^{0.024}_{0.026}$
        \\
        \hline
        $3$
            &-0.010 & ${}^{0.006}_{0.007}$
            &-0.129 & ${}^{0.007}_{0.007}$
            &-0.061 & ${}^{0.007}_{0.009}$
        \\
        \hline
        $4$
            & 0.168 & ${}^{0.009}_{0.008}$
            & 0.220 & ${}^{0.003}_{0.003}$
            & 0.198 & ${}^{0.004}_{0.003}$
        \\
        \hline
        $5$
            &-0.304 & ${}^{0.023}_{0.021}$
            &-0.242 & ${}^{0.007}_{0.008}$
            &-0.294 & ${}^{0.010}_{0.009}$
        \\
        \hline
        $6s$
            &-0.189 & ${}^{0.024}_{0.030}$
            &-0.407 & ${}^{0.014}_{0.013}$
            &-0.346 & ${}^{0.026}_{0.024}$
        \\
        \hline
    \end{tabular}
    \caption{Estimates for the normalized nonvanishing angular observables in
        the SM. The integration ranges are (a) $1\,\GeV^2 \leq q^2 \leq
        6\,\GeV^2$, (b) $14.18\,\GeV^2 \leq q^2 \leq 19.71\,\GeV^2$, and
        $0.02\,\GeV^2 \leq q^2 \leq 19.71\,\GeV^2$.  We normalize the integrated
        angular observables $\langle \Jhat_n\rangle$ to the partially integrated
        decay width $\langle \Gamma\rangle$ for the same integration range.
    }
    \label{tab:jhat-sm}
\end{table}

\begin{table}[t]
    \renewcommand{\arraystretch}{1.3}
    \begin{tabular}{%
        | @{\quad} c @{\quad}%
        | S[table-format=+3.3] @{\,${}^{+}_{-}$\,} S[table-format=1.7] %
        | S[table-format=+3.3] @{\,${}^{+}_{-}$\,} S[table-format=1.7] %
        | S[table-format=+3.3] @{\,${}^{+}_{-}$\,} S[table-format=1.7] %
        |
    }
        \hline
            & \multicolumn{6}{c|}{$\langle \Jhat_n\rangle / \langle\Gamma\rangle$}\\
        $n$
            & \multicolumn{2}{c|}{(a)}
            & \multicolumn{2}{c|}{(b)}
            & \multicolumn{2}{c|}{(c)}
        \\
        \hline
        $1s$
            & 0.132 & ${}^{0.025}_{0.017}$
            & 0.362 & ${}^{0.009}_{0.010}$
            & 0.272 & ${}^{0.021}_{0.021}$
        \\
        \hline
        $1c$
            & 0.574 & ${}^{0.023}_{0.033}$
            & 0.268 & ${}^{0.013}_{0.011}$
            & 0.387 & ${}^{0.028}_{0.028}$
        \\
        \hline
        $2s$
            & 0.044 & ${}^{0.008}_{0.006}$
            & 0.121 & ${}^{0.003}_{0.003}$
            & 0.091 & ${}^{0.007}_{0.007}$
        \\
        \hline
        $2c$
            &-0.574 & ${}^{0.033}_{0.023}$
            &-0.268 & ${}^{0.011}_{0.013}$
            &-0.387 & ${}^{0.028}_{0.028}$
        \\
        \hline
        $3$
            &-0.022 & ${}^{0.013}_{0.009}$
            &-0.151 & ${}^{0.018}_{0.016}$
            &-0.082 & ${}^{0.020}_{0.013}$
        \\
        \hline
        $4$
            & 0.171 & ${}^{0.009}_{0.009}$
            & 0.228 & ${}^{0.007}_{0.008}$
            & 0.207 & ${}^{0.008}_{0.008}$
        \\
        \hline
        $5$
            &-0.271 & ${}^{0.033}_{0.036}$
            &-0.221 & ${}^{0.023}_{0.019}$
            &-0.264 & ${}^{0.021}_{0.029}$
        \\
        \hline
        $6s$
            &-0.172 & ${}^{0.028}_{0.031}$
            &-0.370 & ${}^{0.035}_{0.035}$
            &-0.312 & ${}^{0.031}_{0.041}$
        \\
        \hline
    \end{tabular}
    \caption{Estimates for the nonvanishing angular observables $\Jhat_n$ in
        the SM+SM' basis for real-valued Wilson coefficients. Constraints on
        the Wilson coefficient are taken from data on exclusive semileptonic
        $b\to u$ transitions, see text.  The integration ranges are (a)
        $1\,\GeV^2 \leq q^2 \leq 6\,\GeV^2$, (b) $14.18\,\GeV^2 \leq q^2 \leq
        19.71\,\GeV^2$, and $0.02\,\GeV^2 \leq q^2 \leq 19.71\,\GeV^2$.  We
        normalize the angular observables to the partially integrated decays
        width $\langle \Gamma\rangle$. Note that the quoted sign 
        for the angular observables $\Jhat_5$ and $\Jhat_{6s}$ 
        corresponds to the SM-like solution \eqref{eq:real1} with dominating
        left-handed current. 
        For the solution \eqref{eq:real2}, one simply has to flip the sign 
        of $\Jhat_5$ and $\Jhat_{6s}$.
    }
    \label{tab:jhat-np}
\end{table}

\section{Conclusions}

\label{sec:concl}

The angular analysis of 
exclusive $ \bar B_s \to K^{*+}(\to K\pi) \ell \bar\nu_\ell$  decays 
provides a powerful tool to measure the Cabibbo-Kobayashi-Maskawa (CKM)
element $|V_{ub}|$ in the 
Standard Model (SM) and to constrain new physics (NP) contributions 
to the underlying semileptonic $b \to u \ell \bar\nu_\ell$ transition.
In this article, we have identified relations among the angular observables
that serve as null tests of the SM. 
Furthermore, we have constructed 
optimized observables where, also in the presence of NP,
the dependence on either the hadronic form-factor or the short-distance 
coefficients drops out.
The fact that the same secondary decay, $K^*  \to K\pi$, is used for the
angular analysis of the rare $B \to K^*\ell^+\ell^-$ decay 
can be phenomenologically exploited by measuring certain ratios $R_n$ of 
angular observables from both decays. In the limit where nonfactorizable
effects in $B \to K^*\ell^+\ell^-$ as well as $SU(3)_f$ symmetry
corrections to form-factor ratios can be neglected, the ratios $R_n$ are only
sensitive to short-distance coefficients. In particular, we have shown that
in this way one can directly access the $q^2$-dependence of the 
effective Wilson coefficient function $\wilson[{\rm eff}]{9}(q^2)$ 
in $B \to K^*\ell^+\ell^-$ transitions.

We have combined presently available experimental data on inclusive and
exclusive, leptonic and semileptonic $b \to u$ transitions with theoretical
information on hadronic form factors and decay constants, thereby obtaining
detailed numerical estimates for angular observables and partially integrated
decay widths in $ \bar B_s \to K^{*+}(\to K\pi) \ell \bar\nu_\ell$. Here,
we also allowed for the presence of right-handed currents that could arise
from physics beyond the SM. Using a Bayesian approach for the statistical
treatment of theoretical uncertainties, we have found that -- despite the 
present tensions between different $|V_{ub}|$ determinations -- the SM
is still more efficient in describing the experimental data than its
right-handed extension. In a simultaneous SM fit to $\bar B^0 \to \pi^+ \mu^- \bar \nu_\mu$
(using light-cone sum rule results for low dilepton mass),
$B^- \to \tau^- \bar\nu_\tau$ and $B \to X_u \ell \nu_\ell$ data, we find 
$|V_{ub}|= (4.07 \pm 0.20) \cdot 10^{-3}$ with a p-value of $91\%$. 

On the other hand, right-handed contributions cannot be excluded, either.
In a SM-like scenario with dominating left-handed currents, we found that
the ratio of right-handed over left-handed currets is constrained to $\lesssim 10\%$.
Since the decay rates alone are invariant under parity transformations,
a second solution, with the role of left- and right-handed quark currents
interchanged, is always present.\footnote{Notice that the lepton current with a light SM-like neutrino
is always considered to be left-handed, only.} 
Again, some of the angular observables in $ \bar B_s \to K^{*+}(\to K\pi) \ell \bar\nu_\ell$, \eg the 
leptonic forward-backward asymmetry, are ``parity''-odd and can thus unambigously
test the (dominating) left-handed nature of semileptonic 
$b \to u$ currents. In this case, one would obtain strong 
constraints on the flavour sector of NP models with generic right-handed currents.
(For a recent attempt to construct a left-right symmetric NP model based on the
Pati-Salam gauge group, which can accomodate naturally
small right-handed $b\to u$ currents, see \cite{Feldmann:2015unp}.)

A crucial ingredient of our analysis has been the implementation of
hadronic uncertainties. Improvements of our theoretical understanding
of nonperturbative QCD effects (see also notes added below) 
would lead to more stringent constraints on the value of $|V_{ub}|$ and 
the possible size of right-handed $b \to u$ currents. 
In particular, predictions from lattice or light-cone sum rules 
for form-factor ratios with $\bar{B}$ and $\bar{B}_s$ initial
states (including correlations between input parameters),
and similarly between $B \to \pi$ form factors and the $B$-meson
decay constant, would be helpful in this respect.

\subsection*{Notes added:}

In the final phase of this work, the LHCb collaboration
measured the ratio of the exclusive semileptonic branching fractions of
$\Lambda_b\to p\mu^-\bar\nu_\mu$ and $\Lambda_b \to \Lambda_c \mu^-\bar\nu_\mu$
\cite{Sutcliffe:2002528,Owen:2002890}.  Assuming SM-like $b\to c\mu^-\bar\nu_\mu$
transitions, with knowledge of the magnitude of $|V_{cb}|$ and using
information on the relevant form factors \cite{Detmold:2015aaa}, this ratio can
be used to extract the branching fraction $\mathcal{B}(\Lambda_b \to p \mu^-
\bar\nu_\mu)$. As such, the branching fraction is a very powerful new
constraint.  However, in light of the present tension in the determination of
$V_{cb}$ from both inclusive and exclusive $b\to c\ell\bar\nu_\ell$ decays, and in order to 
follow the logical line of this article, the new LHCb measurement should only be used 
in a setup that accounts for NP in both $b \to u$ and $b\to c$ transitions.

Another article \cite{Straub:2015ica} that was published in the final phase of this work provides updated
LCSR results for the hadronic form factors for $\bar{B}_s \to K^*$ transitions,
which include correlation information among the form factors. This development will help
to further reduce theory uncertainties for this decay.

In recent lattice studies of the $B \to \pi$ form factors \cite{Flynn:2015mha,Lattice:2015tia}, also the correlation matrix between
the relevant hadronic fit parameters has been provided. This will also allow to include the high-$q^2$ data
for the $\bar B \to \pi \ell \bar\nu_\ell$ decay in our statistical procedure, which could and should
be used in future updates of our results.

\section*{Acknowledgment}

D.v.D. acknowledges time and effort spent by Martin Jung on checking the
angular distribution in the early phases of this work, and also for the initial
idea to investigate the full angular distribution of $B\to V\ell \nu_\ell$ decays
for the complete basis of dimension-six operators.
This work is supported in parts by the Bundesministerium f\"ur Bildung und
Forschung (BMBF), and the Deutsche Forschungsgemeinschaft (DFG) within research
unit FOR 1873 (``QFET'').

\appendix

\section{Form Factors}
\label{app:ff}

There are in general 7 independent hadronic form factors for $B_s \to K^*$ transitions.
Commonly, these are denoted as $V$, $A_{0,1,2}$, $T_{1,2,3}$, see \eg the definition
in \cite{Ball:2004rg}.
For our purpose, it is more convenient 
to start with a definition of form factors in a helicity basis,
\begin{equation}
\begin{aligned}
    F_\pm      & \equiv \frac{i}{M_{B_s}}
                 \langle K^*(k, \eta)|\bar{u} \slashed{\eps}^*_{\pm}(1 - \gamma_5)b|\bar{B}_s(p)\rangle\,,\cr 
    F_{0}      & \equiv \frac{-i \sqrt{q^2}}{M^2_{B_s}}
                 \langle K^*(k, \eta)|\bar{u} \slashed{\eps}^*_{0}(1 - \gamma_5)b|\bar{B}_s(p)\rangle\,,\cr 
    F_{t}      & \equiv \frac{i \sqrt{q^2}}{M^2_{B_s}}
                 \langle K^*(k, \eta)|\bar{u} \slashed{\eps}^*_{t}(1 - \gamma_5)b|\bar{B}_s(p)\rangle\,,
\end{aligned}
\end{equation}
and
\begin{equation}
\begin{aligned}
    F^T_\pm    & \equiv \frac{1}{M^2_{B_s}}
                 \langle K^*(k, \eta)|\bar{u} \sigma_{\mu\nu} \epsilon^{\mu*}_\pm q^\nu (1 + \gamma_5) b|\bar{B}_s(p)\rangle\,,\cr 
    F^T_{0}    & \equiv \frac{1}{M_{B_s} \sqrt{q^2}} \langle K^*(k, \eta)|\bar{u}  \sigma_{\mu\nu} \epsilon^{\mu*}_0 q^\nu (1 + \gamma_5) b|\bar{B}_s(p)\rangle\,,\\
\end{aligned}
\end{equation}
which is related to the one proposed in \cite{Bharucha:2010im}. 
However, compared to \cite{Bharucha:2010im}, we have chosen a normalization convention
such that all form factors are  finite in the limit $q^2 \to t_- \equiv (M_{B_s} - M_{K^*})^2$, 
and nonzero in the limit $q^2 \to 0$.  
In the above definition, $\eta$ denotes the physical polarization of the
$K^*$ meson, and $\epsilon$ stands for an auxiliary polarization vector 
of the dilepton system with polarization states $t,\pm 1,0$.
Notice, that the form factor for the pseudoscalar current
is not independent, but from the equations of motion can be related to $F_t$,
\begin{equation}
    \langle K^*(k,\eta)|\bar{u} \gamma_5 b|\bar{B}_s\rangle = -i \frac{M_{B_s}^2}{m_b + m_u} F_t\,.
\end{equation}

Instead of the helicity form factors $F_\pm$, we will use the linear
combinations
\begin{equation}
\begin{aligned}
    F_{\para(\perp)}   & \equiv \frac{1}{\sqrt{2}}(F_- \pm F_+)\,, &
    F^T_{\para(\perp)} & \equiv \frac{1}{\sqrt{2}}(F^T_- \pm F^T_+)\,, &
\end{aligned}
\end{equation}
which simplify the analytical expressions for the angular observables.
The explicit relations between our and the traditional form factor basis
read 
\begin{equation}
\begin{aligned}
    F_\perp & = \frac{\sqrt{2 \lambda}}{M_{B_s}(M_{B_s} + M_{K^*})} V\,,
\end{aligned}
\end{equation}
for the vector form factor,
\begin{equation}
\begin{aligned}
    F_\para & = \sqrt{2} \, \frac{M_{B_s} + M_{K^*}}{M_{B_s}} A_1\,,\cr 
    F_0     & = \frac{(M_{B_s} + M_{K^*})^2 \, (M_{B_s}^2 - M_{K^*}^2 - q^2) \, A_1 - \lambda \, A_2}{2 M_{K^*} \, M^2_{B_s} \, (M_{B_s} + M_{K^*})}\cr 
            & = \frac{8 M_{K^*} \, A_{12}}{M_{B_s}}\,,\cr 
    F_t     & = \frac{\sqrt{\lambda}}{M_{B_s}^2} \, A_0\,,
\end{aligned}
\end{equation}
for the  axialvector currents, and
\begin{equation}
\begin{aligned}
    F^T_\perp & = \frac{\sqrt{2 \lambda}}{M_{B_s}^2} \, T_1\,,\cr 
    F^T_\para & = \frac{\sqrt{2} \, (M_{B_s}^2 - M_{K^*}^2)}{M_{B_s}^2} \, T_2\,,\cr 
    F^T_0     & = \frac{(M_{B_s}^2 - M_{K^*}^2) \, (M_{B_s}^2 + 3 M_{K^*}^2 - q^2) \, T_2 - \lambda \, T_3}{2 M_{K^*} \, M_{B_s} \, (M_{B_s}^2 - M_{K^*}^2)}\cr 
              & = \frac{4 M_{K^*} \, T_{23}}{M_{B_s} + M_{K^*}}\,,
\end{aligned}
\end{equation}
for the tensor current. In the above equations, the form factors $A_{12}$ and $T_{23}$
are defined as in \cite{Horgan:2013hoa}.

\begin{table}[pt]
\renewcommand{\arraystretch}{1.3}
\begin{tabular}{|c|ccc|}
    \hline
    $q^2$ [GeV$^2$]  & $0$               & $15.00$           & $19.21$           \\
    \hline
    $V(q^2)$         & $0.311 \pm 0.026$ & $0.872 \pm 0.066$ & $1.722 \pm 0.062$ \\
    $A_1(q^2)$       & $0.233 \pm 0.023$ & $0.427 \pm 0.015$ & $0.548 \pm 0.015$ \\
    $A_2(q^2)$       & $0.181 \pm 0.025$ & --                & --                \\
    $A_{12}(q^2)$    & --                & $0.342 \pm 0.016$ & $0.408 \pm 0.016$ \\
    \hline
\end{tabular}\\[1em]
\begin{tabular}{|c|cc|cc|cc|}
    \hline
                     & \multicolumn{2}{c|}{$V$} & \multicolumn{2}{c|}{$A_1$} & \multicolumn{2}{c|}{$A_{12}$}  \\
    $q^2$ [GeV$^2$]  & $15.00$ & $19.21$        & $15.00$ & $19.21$          & $15.00$ & $19.21$           \\
    \hline
    $15.00$          & $1.000$ & $0.271$        & $1.000$ & $0.305$          & $1.000$ & $0.334$           \\
    $19.21$          & --      & $1.000$        & --      & $1.000$          & --      & $1.000$           \\
    \hline
\end{tabular}
    \caption{Theory inputs for the $B_s \to K^*$ form factor fits.
    Top: Form factor values at $q^2=0$ are taken from LCSR calculations in \cite{Ball:2004rg};
    values at $q^2=15~\GeV^2$ and $q^2=19.21~\GeV^2$ are taken from lattice QCD simulations \cite{Horgan:2013hoa}.
    Bottom: Correlation information for the lattice QCD inputs.
    The lattice QCD values and correlations are produced from the joint PDF given in
    \cite[Table XXIX]{Horgan:2013hoa} using $5\cdot 10^5$ samples.
    \label{tab:ff-constraints}}
\end{table}
The form factors fulfill endpoint relations \cite{Bharucha:2010im,Hiller:2013cza}\footnote{Note that
the endpoint relation for the $\perp$ form factor in \cite[appendix B]{Bharucha:2010im}
should read $\lim_{q^2 \to t_-} B_{V,1} / B_{V,2} = 0$.} which in our convention read
\begin{equation}
\label{eq:ep-relations}
\begin{aligned}
    \lim_{q^2 \to t_-} F_\perp             & = \lim_{q^2 \to t_-} F_t = 0\,, \cr 
    \lim_{q^2 \to t_-} \frac{F_\para}{F_0} & = \frac{\sqrt{2} M_{B_s}}{M_{B_s} - M_{K^*}}\,,
\end{aligned}
\end{equation}
with $t_\pm \equiv (M_{B_s} \pm M_{K^*})^2$. We will use these relations for 
our form-factor parametrization in the numerical fit.
To this end, we consider a modified ``$z$-expansion''
and write
\begin{align}
\label{eq:ff-param}
    F_\perp(q^2)     & = \frac{\sqrt{\lambda}}{M_{B_s}^2 - M_{K^*}^2} \, P(q^2, M_{B^*}^2) \, F_\perp(0)\cr
                     & \quad \times \left[1 + b_\perp \left(z(q^2, t_0) - z(0, t_0)\right)\right]\,,\cr 
    F_{\para,0}(q^2) & = P(q^2, M_{B_1}^2) \, F_{\para,0}(0)\cr 
                     & \quad \times \left[1 + b_{\para,0} \left(z(q^2, t_0) - z(0, t_0)\right)\right]\,,\cr
    F_t(q^2)         & = \frac{\sqrt{\lambda}}{M_{B_s}^2 - M_{K^*}^2} \, P(q^2, M_{B}^2) \, F_t(0)\cr 
                     & \quad \times \left[1 + b_t \left(z(q^2, t_0) - z(0, t_0)\right)\right]\,.
\end{align}
Here, the prefactors contain global kinematic factors, the form-factor normalization
at $q^2=0$, together with the leading pole behaviour from the lowest resonances above the 
semileptonic decay region, $P(q^2, M^2)^{-1}
\equiv 1 - q^2/M^2$. The remaining $q^2$-dependence for each form factors 
is parametrized by a shape parameter $b_i$.
The variable $z(q^2,t_0)$ is obtained from the conformal mapping, (see \eg \cite{Boyd:1994tt,Becher:2005bg,Bourrely:2008za})
\begin{equation}
    z(a, b) \equiv \frac{\sqrt{t_+ - a} - \sqrt{t_+ - b}}{\sqrt{t_+ - a} + \sqrt{t_+ - b}}\,.
\end{equation} 
Here we choose $t_0 = t_+ - \sqrt{t_+  (t_+ - t_-)}$ which minimizes $|z|$ in the decay region.
For the resonance masses we use $M_B = 5279$~\MeV, $M_{B^*} = 5325~\MeV$ and $M_{B_1} = 5724~\MeV$
\cite{Agashe:2014kda}.  The above parametrization \refeq{ff-param}
automatically fulfills the end-point relation \refeq{ep-relations} for
$F_\perp$. The end-point relation for $F_\para/F_0$ is fulfilled by imposing
\begin{widetext}
\begin{equation}
    b_0 \equiv \frac{1}{z(0, t_0) - z(t_-, t_0)}\left(1 - \frac{F_\para(0)}{F_0(0)} \sqrt{\frac{t_-}{2 M_{B_s}^2}} \left[1 + b_\para(z(t_-,t_0) - z(0,t_0))\right]\right)\,.
\end{equation}
\end{widetext}

We fit the $B_s\to K^*$ helicity form factors $F_{\perp,\para,0}$ to the nine
constraints listed in \reftab{ff-constraints}. Our fit uses five parameters,
\begin{equation}
    \vec\theta_\text{FF} = \left(F_\perp(0), F_\para(0), F_0(0), b_\perp, b_\para\right)
\end{equation}
which represent the three normalizations $F_{\perp,\para,0}(q^2 = 0)$, as well
as two independent shape parameters $b_{\perp,\para}$. As \emph{a-priori}
probability $P_0(\vec\theta_{FF})$ we choose uncorrelated uniform distributions
with a generous support (to be compared with \eqref{eq:Fresults} below),
\begin{equation}
      0  \leq F_{\perp,\para,0}(0) \leq 1\,, \quad 
    -10  \leq b_\perp \leq 0\,, \
     -5  \leq b_\para \leq +5\,.
\end{equation}
The likelihood $P(\text{theory} | \vec\theta_\text{FF})$ is constructed as the
product of uncorrelated Gaussian likelihoods for each of the LCSR results for
the form factors $V$, $A_1$ and $A_2$, as well as the joint multivariate
Gaussian likelihood for the lattice QCD results.  All of these are listed in
\reftab{ff-constraints}.\\

The \emph{a-posteriori} PDF is obtained as usual via Bayes' theorem,
\begin{equation}
\label{eq:posterior-ff}
    P(\vec\theta_\text{FF} | \text{theory}) = \frac{P(\text{theory} | \vec\theta_\text{FF}) P_0(\vec\theta_\text{FF})}{\int \dd\vec\theta_\text{FF}\, P(\text{theory} | \vec\theta_\text{FF}) P_0(\vec\theta_\text{FF})}\,.
\end{equation}
For all applications here and in \refsec{numerics}, we draw $10^6$ samples
from the \emph{a-posteriori} distribution.

The best-fit point, and the 1D-marginalized minimal intervals at $68\%$ probability are found to be
\begin{equation}
\label{eq:Fresults}
\begin{aligned}
    F_\perp(0) & =  0.349 \pm 0.037\,, &
    b_\perp    & = -4.9^{+1.0}_{-1.1}\,, \\
    F_\para(0) & =  0.379 \pm 0.031\,, &
    b_\para    & = +0.07 \pm 0.40\,. \\
    F_0(0)     & =  0.314 \pm 0.041\,, 
\end{aligned}
\end{equation}
Although the 1D-marginalized distributions are symmetric and resemble Gaussian
distributions, we find that the distribution in
\refeq{posterior-ff} is distinctly non-Gaussian. 
We therefore use the posterior samples to carry out the uncertainty
propagation.

\section{$\bar{B}_s \to K^*(\to K \pi)\ell^-\bar\nu_\ell$ Decay Amplitude}
\label{app:hme}

In this appendix we give details on the parametrization of the matrix element
for the decay $\bar{B}_s\to K^{*+}\ell^-\bar\nu_\ell$, with the subsequent
decay $K^{*+}\to (K\pi)^+$. We decompose the matrix element as in \cite{Bobeth:2012vn}
\begin{multline}
    \mathcal{M} = \mathcal{F} \big\{X_S \[\bar\ell \nu\] + X_P \[\bar\ell\gamma_5\nu\]\\
                + X_V^\mu \[\bar\ell \gamma_\mu\nu\] + X_A^\mu \[\bar\ell\gamma_\mu\gamma_5\nu\]
                + X_T^{\mu\nu} \[\bar\ell \sigma_{\mu\nu}\nu\]\big\}
\end{multline}
with the prefactor
\begin{equation}
    \mathcal{F} = i \sqrt{2} \, G_{\rm F}\, V_{ub} \, g_{K*K\pi} \, D_{K^*} \, \krf\,,
\end{equation}
and $\krf \equiv \sqrt{\lambda(M_{K^*}^2, M_K^2, M_\pi^2)} / 2 M_{K^*}$.
In the small-width approximation we replace the $K^*$ resonance by
\begin{equation}
\begin{aligned}
    |D_{K^*}(k^2)|^2
        & \simeq \frac{1}{(k^2 - M_{K^*}^2)^2 + M_{K^*}^2 \Gamma_{K^*}^2}\\
        & \to \frac{\pi}{M_{K^*} \Gamma_{K^*}}\delta(k^2 - M_{K^*}^2)
\end{aligned}
\end{equation}
where $\Gamma_{K^*}$ denotes the total decay width of the $K^*$ meson.
Since $\Gamma_{K^*} \simeq \Gamma[K^* \to K\pi]$ to very good approximation, we use
\begin{equation}
    \Gamma_{K^*} = \frac{|g_{K^*\to K\pi}|^2 \krf^3}{48 \pi M_{K^*}^5}\,.
\end{equation}\\

Our parametrization of the hadronic matrix element of $B\to V(\to P_1 P_2) \ell^-\bar\nu_\ell$ decays
differes from the one in \cite{Bobeth:2012vn} due to different conventions for the Levi-Civita tensor,
the phase convention for the polarization vectors, and the fact that in this decay only left-handed
lepton currents contribute. We use
\begin{align}
  N X_S & = \frac{i}{4}\cos\theta_V A_t^L = -N X_P\,,
\end{align}
and
\begin{align}
  N X_V^\mu &= -N X_A^\mu \cr 
      & = +\frac{i}{4}\cos\theta_V \eps^\mu(0) A_0^L\cr
      & +\frac{i}{8}\sin\theta_V \eps^\mu(+) e^{+i\phi}\[A_\perp^L + A_\para^L\] \cr 
      & +\frac{i}{8}\sin\theta_V \eps^\mu(-) e^{-i\phi}\[A_\perp^L - A_\para^L\]  \,, \\
\end{align}
and
\begin{align}
  N X_T^{\mu\nu}
      & = \cos\theta_V \eps^\mu(+)\eps^\nu(-) A_{\parallel\perp} \cr 
      & + \frac{\sin\theta_V}{\sqrt{2}} \eps^\mu(t)\eps^\nu(+)e^{+i\phi}A_{t\perp} \cr
      & + \frac{\sin\theta_V}{\sqrt{2}} \eps^\mu(t)\eps^\nu(-)e^{-i\phi}A_{t\perp} \cr
      & - \frac{\sin\theta_V}{\sqrt{2}} \eps^\mu(0)\eps^\nu(+)e^{+i\phi}A_{0\para} \cr
      & - \frac{\sin\theta_V}{\sqrt{2}} \eps^\mu(0)\eps^\nu(-)e^{-i\phi}A_{0\para} \,.
\end{align}
Using the normalisation constant $N$ as given in \refeq{normalisation} and the 
general operator basis (\ref{eq:op-basis}), we obtain for the individual amplitude
contributions
\begin{align}
    A_0^L          & = -4 N \, \frac{M_{B_s}^2}{\sqrt{q^2}} \, 
              (\wilson{V,L} - \wilson{V,R}) \, F_0(q^2)\,,\cr 
    A_\perp^L      & = +4 N \, M_{B_s} \, (\wilson{V,L} + \wilson{V,R}) \, F_\perp(q^2)\,,\cr 
    A_\para^L      & = -4 N \, M_{B_s} \, (\wilson{V,L} - \wilson{V,R}) \, F_\para(q^2)\,,\cr 
    A_t^L          & = -4 N \Big[\frac{m_\ell \, M_{B_s}}{q^2}\big(\wilson{V,L} - \wilson{V,R}\big)\cr 
          & \quad + \frac{M_{B_s}^2}{m_b}\big(\wilson{S,L} - \wilson{S,R}\big)\Big] F_t(q^2)\,,
\end{align}
and
\begin{align}
    A_{\para\perp} & = +8 N \, M_{B_s} \, \wilson{T} \, F^T_0(q^2)\,,\cr 
    A_{t\perp}     & = 4\sqrt{2} N \, \frac{M_{B_s}^2}{\sqrt{q^2}} \, \wilson{T} \, F^T_\perp(q^2)\,,\cr 
    A_{0\para}     & = 4\sqrt{2} N \, \frac{M_{B_s}^2}{\sqrt{q^2}} \, \wilson{T} \, F^T_\para(q^2)\,.
\end{align}


\section{Angular Observables for $B\to V\ell\nu_\ell$}
\label{app:ang-dist-full}

In the limit $m_\ell \to 0$, the angular observables $\hat J_n$ read
\begin{align}
    \hat J_{1s}  & 
      = \frac{3}{16} \, \[3 |A_\perp^L|^2 + 3 |A_\para^L|^2 + 16 |A_{0\para}|^2 + 16 |A_{t\perp}|^2\]\,,\cr 
    \hat J_{1c}  & = \frac{3}{4} \, \[|A_0^L|^2 + 2 |A_t^L|^2 + 8 |A_{\para\perp}|^2\]\,,\cr
    \hat J_{2s}  & = \frac{3}{16}\, \[|A_\perp^L|^2 + |A_\para^L|^2 - 16 |A_{0\para}|^2 - 16 |A_{t\perp}|^2\]\,,\cr
   \hat J_{2c}  & = -\frac{3}{4}\, \[|A_0^L|^2 - 8 |A_{\para\perp}|^2\]\,,\cr
    \hat J_3     & = \frac{3}{8}\, \[|A_\perp^L|^2 - |A_\para^L|^2 + 16 |A_{0\para}|^2 - 16 |A_{t\perp}|^2\]\,,\cr
    \hat J_4     & = \frac{3}{4\sqrt{2}}\, \Re{A_0^L A_\para^{L*} - 8\sqrt{2} A_{\para\perp} A_{0\para}^*}\,,
\end{align}
and 
\begin{align}
\hat J_5     & = \frac{3}{2\sqrt{2}} \, \Re{A_0^L A_\perp^L + 2\sqrt{2} A_{0\para} A_t^{L*}}\,,\cr
    \hat J_{6s}  & = \frac{3}{2}\, \Re{A_\para^L A_\perp^{L*}}\,,\cr
   \hat J_{6c}  & = -6 \, \Re{A_{\para\perp} A_t^{L*}}\,,\cr
    \hat J_7     & = \frac{3}{2\sqrt{2}} \, \Im{A_0^L A_\para^{L*} - 2\sqrt{2} A_{t\perp} A_t^{L*}}\,,\cr
    \hat J_8     & = \frac{3}{4\sqrt{2}}\, \Im{A_0^L A_\perp^{L*}}\,,\cr
    \hat J_9     & = \frac{3}{4}\,\Im{A_\perp^L A_\para^{L*}}\,.
\end{align}


\bibliographystyle{apsrev4-1}
\bibliography{references.bib}

\end{document}